\documentclass
[prb,twocolumn,amsmath,amssymb,amsfonts,unsortedaddress]{revtex4}%
\usepackage[dvips]{graphicx}
\usepackage{psfrag}
\usepackage{dcolumn}
\usepackage{bm}
\usepackage{amsmath}
\usepackage{amsfonts}
\usepackage{amssymb}%
\allowdisplaybreaks[1]
\begin{document}
\title{\indent Solution of the Multi-Channel Anderson Impurity Model\\--~Ground State and Thermodynamics~--}
\author{C.~J.~Bolech$^{1,2}$}
\affiliation{$^{1}$ D\'{e}partement de Physique de la Mati\`{e}re Condens\'{e}e,
Universit\'{e} de Gen\`{e}ve, Quai Ernest Ansermet 24, CH-1211 Gen\`{e}ve 4, Switzerland}
\author{N.~Andrei$^{2}$}
\affiliation{$^{2}$ Center for Materials Theory, Serin Physics Laboratory, Rutgers
University, 136 Frelinghuysen Road, Piscataway, New Jersey 08854-8019, USA}
\date{July 14$^{\text{th}}$, 2004}

\begin{abstract}
We present the solution of the $SU\!\left(  N\right)  \otimes SU\!\left(
M\right)  $ Anderson impurity model using the Bethe-Ansatz. We first explain
what extensions to the formalism were required for the solution. Subsequently
we determine the ground state and derive the thermodynamics over the full
range of temperature and fields. We identify the different regimes of valence
fluctuation at high temperatures, followed by moment formation or intrinsic
mixed valence at intermediate temperatures and a low temperature non-Fermi
liquid phase. Among other things we obtain the impurity entropy, charge
valence, and specific heat over the full range of temperature. We show that
the low-energy physics is governed by a line of fixed points. This describes
non-Fermi-liquid behavior in the integral valence regime, associated with
moment formation, as well as in the mixed valence regime where no moment forms.

\end{abstract}
\pacs{}
\maketitle

\section{Introduction}

Heavy-fermion materials have been a source of interest and puzzlement since
the experimental discovery of superconductivity in \textrm{CeCu}$_{2}%
$\textrm{Si}$_{2}$ in 1979.\cite{sablmfs79} By the turn of the 1990s the
non-Fermi liquid character of these materials was coming to the center of
attention (for a recent review see Ref.~%
[\onlinecite{stewart01}]%
), coincidentally with the interest on marginal Fermi liquids generated by the
normal state of high-$T_{\mathrm{c}}$ superconductors. Since then, the number
of examples of violation of Landau's Fermi liquid theory among lantanides and
actinides has multiplied.

Current theories trying to explain the non-Fermi liquid behavior in
\textrm{d}- and \textrm{f}-electron metals can be classified into three broad
categories: (i) models based on multi-channel Kondo physics, (ii) models
considering the proximity of a quantum critical point, and (iii) models based
on single-channel Kondo physics but in the presence of disorder that induces a
distribution of impurity energy scales. These three ingredients are not
mutually exclusive and a number of recent theories try to address their
effects in different combinations (see
[\onlinecite{stewart01}]
for references). In this article we will be concerned with the first class of
models. This approach originated with the work of Cox,\cite{cox87} in turn
motivated by the unusual experimental results in the heavy-fermion compound
\textrm{UBe}$_{13}$.\cite{sfw83} He argued that the notably weak magnetic
field dependence of the specific heat of this material excludes the usual
magnetic Kondo effect and proposed instead that the observed anomalous
behavior derives from the quenching of quadrupolar degrees of freedom. In this
case, the spin of the conduction band electrons plays the role of a channel
degree of freedom. This is the two-channel quadrupolar Anderson model,
describing tetravalent uranium impurities in a cubic-symmetric metallic
matrix. The model was later generalized to include other crystal symmetries as
well as more complicated crystal field splittings; for a review see Ref.~%
[\onlinecite{cox98}]%
. In more detail, Hund's rules and spin-orbit coupling in the presence of a
cubic crystalline electric field lead to the modeling of a \textrm{U} ion in a
\textrm{Be}$_{13}$ host by a $\Gamma_{6}$ Kramers doublet in a $5$%
\textrm{f}$^{3}$ configuration and a quadrupolar (nonmagnetic) doublet
$\Gamma_{3}$ in the $5$\textrm{f}$^{2}$ configuration. The doublets hybridize
with conduction electrons in a $\Gamma_{8}$ representation carrying both spin
and quadrupolar quantum numbers. This single-impurity approach was not
uncontroversial.\cite{at91,ramirez94} For instance, whether the energy
splitting between the two doublets is sufficiently large for a quadrupolar
Kondo scenario to be viable is still largely unresolved. Aliev \textit{et al.}
presented experimental evidence suggesting that both doublets may in fact be
nearly degenerate, pointing to a mixed-valent state with a novel type of
interplay between magnetic and quadrupolar two-channel type
screening.\cite{aliev95}

In this article we will study the general multi-channel Anderson impurity
model that includes as particular cases the two-channel model and, to some
extent, most of its generalizations alluded to above.\cite{cox98} The
so-called $SU\!\left(  N\right)  \otimes SU\!\left(  M\right)  $ Anderson
impurity model, in its pseudo-particle representation, is given by the
following Hamiltonian:%

\begin{multline*}
H_{\mathrm{MchA}}=H_{\text{\textrm{host}}}+\varepsilon_{q}\sum_{\alpha}%
b_{\bar{\alpha}}^{\dagger}b_{\bar{\alpha}}+\varepsilon_{s}\sum_{\sigma
}f_{\sigma}^{\dagger}f_{\sigma}+\\
+V\sum_{\bar{\alpha},\sigma}\left[  f_{\sigma}^{\dagger}b_{\bar{\alpha}}%
\psi_{\alpha,\sigma}\left(  0\right)  +\psi_{\alpha,\sigma}^{\dagger}\left(
0\right)  b_{\bar{\alpha}}^{\dagger}f_{\sigma}\right]
\end{multline*}
subject to the constraint: $\sum_{\alpha}b_{\bar{\alpha}}^{\dagger}%
b_{\bar{\alpha}}+\sum_{\sigma}f_{\sigma}^{\dagger}f_{\sigma}=1$. The first
term in the Hamiltonian describes the host in which the impurity is embedded.
For our purposes, we model it as a linearized Fermi band,%
\[
H_{\text{\textrm{host}}}=\sum_{\alpha,\sigma}\int\psi_{\alpha,\sigma}%
^{\dagger}\left(  x\right)  \left(  -i\partial_{x}\right)  \psi_{\alpha
,\sigma}\left(  x\right)  ~dx~\text{.}%
\]
The second and third terms model two multiplets with energies $\varepsilon
_{s}$ and $\varepsilon_{q}$ and quantum numbers $\sigma\in SU\!\left(
N\right)  $ and $\bar{\alpha}\in SU\!\left(  M\right)  $,
respectively.\footnote{Indices $\alpha$ and $\sigma$ transform according to
the corresponding fundamental representation, the notation $\bar{\alpha}$
indicates that the index transforms instead with the complex conjugate
representation.} We will refer to this two quantum numbers as generalized spin
and flavor --~in reference to the Kramers (magnetic) and non-Kramers
(quadrupolar) doublets of the two-channel case. The last term in the
Hamiltonian describes the hybridization of the host electrons with the impurity.%

\begin{figure}
[tb]
\begin{center}
\includegraphics[
height=1.2194in,
width=3.2258in
]%
{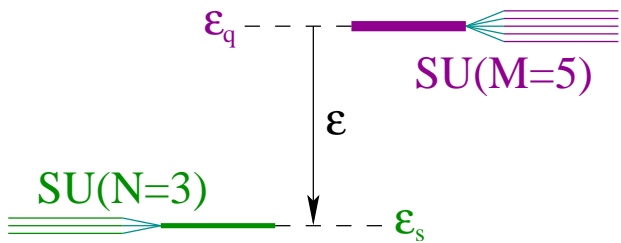}%
\caption{Impurity level scheme for the case of an Anderson model with a
three-fold degenerate magnetic configuration of energy $\varepsilon_{s}$ and a
five-fold degenerate quadrupolar one of energy $\varepsilon_{q}$. The energy
difference $\varepsilon=\varepsilon_{s}-\varepsilon_{q}$ is indicated with an
arrow.}%
\label{levels}%
\end{center}
\end{figure}

An illustrative level scheme is given in Fig.~\ref{levels} where the magnetic
configuration is taken to lie lower in energy that the quadrupolar one. The
particular example is for an Anderson model with $SU\!\left(  N=3\right)
\otimes SU\!\left(  M=5\right)  $ symmetry. The impurity can switch between
the two configurations by giving or taking an electron from the host band, a
process that takes place with an overlap amplitude given by $V$.

As a result of the constraint, the impurity Hilbert space is restricted to
$N+M$ states. Namely, $N$ flavorless spin states $\left\vert \sigma
\right\rangle \equiv f_{\sigma}^{\dagger}\left\vert 0\right\rangle $ plus $M$
spinless flavor states $\left\vert \bar{\alpha}\right\rangle \equiv
b_{\bar{\alpha}}^{\dagger}\left\vert 0\right\rangle $. Defining these states
explicitly, we can rewrite the Hamiltonian in a different notation that
automatically accounts for the Hilbert space restriction,
\begin{multline*}
H_{\mathrm{MchA}}=H_{\text{\textrm{host}}}+\varepsilon_{q}\sum_{\alpha
}\left\vert \bar{\alpha}\right\rangle \left\langle \bar{\alpha}\right\vert
+\varepsilon_{s}\sum_{\sigma}\left\vert \sigma\right\rangle \left\langle
\sigma\right\vert +\\
+V\sum_{\alpha,\sigma}\left[  \left\vert \sigma\right\rangle \left\langle
\bar{\alpha}\right\vert \psi_{\alpha,\sigma}\left(  0\right)  +\psi
_{\alpha,\sigma}^{\dagger}\left(  0\right)  \left\vert \bar{\alpha
}\right\rangle \left\langle \sigma\right\vert \right]  .
\end{multline*}

Both forms of the Hamiltonian are completely equivalent when the constraint is
treated exactly and both are widely used in the literature.\cite{cox98} It is
the constraint acting in the Hilbert space that plays the role of a strong
interaction and renders the problem highly non-perturbative.

As a side remark notice that if either $M$ or $N$ is put to one, the model
reduces to a degenerate single-channel Anderson model in the limit of infinite
Coulomb repulsion (a magnetic or a quadrupolar version of it,
respectively).\cite{schlottmann89,Hewson} While the standard single-channel
$SU\!\left(  2\right)  $ Anderson model\cite{anderson61} was found to be
integrable,\cite{wiegmann80,ko81} its $SU\!\left(  N\right)  $ generalization
is not integrable except for the strong repulsion limit when the impurity is
constrained not to exceed single
occupancy.\cite{schlottmann83a,schlottmann83b}

Even though the multi-channel Anderson model was put forward more than fifteen
years ago, progress in its theoretical understanding has been slow. In the
two-channel case ($N=M=2$) most of the early knowledge of its unusual physics
came from the integer valence limit. In this limit the model maps onto the
two-channel Kondo model for which the Bethe-Ansatz solution was
available.\cite{ad84,tsvelik85,ss93} Also Numerical Renormalization Group
(NRG)\cite{cln80,pc91} and Boundary Conformal Field Theory (BCFT)\cite{al91}
studies were carried out. Only more recently some progress was made in the
study of the mixed valence regime of the two-channel Anderson model using
Monte Carlo\cite{sac98} and NRG\cite{kc99,anders04} methods. The more general
multi-channel case is, however, not quite within the present reach of NRG and
other approaches. On the other hand, the large $N$ and $M$ case constitutes
the natural starting point for alternative approaches like the Non-Crossing
Approximation (NCA),\cite{cr93,sac98} the \textit{conserving slave boson
theory},\cite{kwc97,kw98} or other types of $1/N$-expansions.\cite{tomk97}
Since the general multi-channel Anderson model can be regarded as the
extension of the infinitely-repulsive degenerate single-channel Anderson model
to the multi-channel case, the question about its integrability arises
naturally. During the last couple of years the integrability was established,
opening up the possibility of a full understanding of the model. In previous
work, we presented the Bethe-Ansatz solution for the two-channel
case.\cite{ba02} Subsequently, we developed the critical low-energy theory of
that model using BCFT and combining it with results from Thermodynamic
Bethe-Ansatz.\cite{jab03,jab04}

\subsection{Preview of Main Results}

In the present work we will give a detailed and, to a large extent,
self-contained account of the Bethe-Ansatz solution of the general
multi-channel Anderson model. We shall show that, as in the case of the two
channel model, the low energy physics of the impurity is governed by a line of
boundary fixed points with a non-trivial residual impurity entropy that is
constant along the line. We shall identify two energy scales ($T_{\mathrm{H}}$
and $T_{\mathrm{L}}$) that govern the screening process of the impurity
degrees of freedom. The screening occurs in two stages parameterized by these
scales as will be seen, for instance, in the temperature dependence of the
impurity entropy. The mixed valence regime will be discussed in detail,
stressing not only the differences, but also the unexpected similarities with
the integer valence cases.%

\begin{figure}
[th]
\begin{center}
\includegraphics[
height=2.156in,
width=3.3702in
]%
{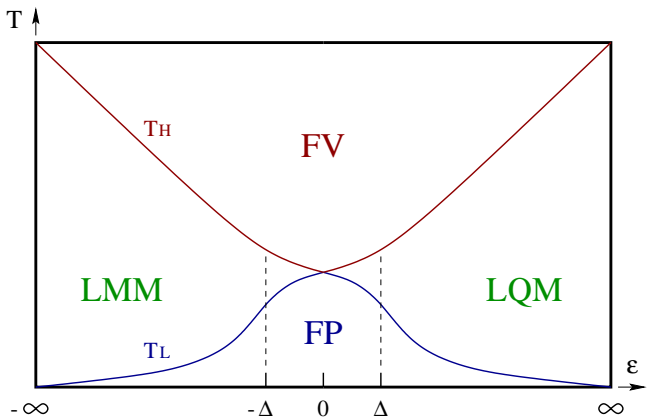}%
\caption{Schematic representation of the two temperature scales indicating the
crossovers among different regimes: fluctuating valence (FV), fixed point
(FP), and local magnetic and quadrupolar moment regimes (LMM and LQM).}%
\label{teps}%
\end{center}
\end{figure}

In Fig.~\ref{teps} we present the picture that emerges for the different
regimes of the model. As a function of temperature and energy difference
between impurity configurations ($\varepsilon=\varepsilon_{s}-\varepsilon_{q}%
$), we will characterize the different regimes (FV, LMM and LQM, and FP; see
figure caption). In particular, moment formation takes place (for a fixed and
suitably large $\varepsilon$) as the temperature falls below $T_{\mathrm{H}}$;
the moment being magnetic or quadrupolar depending on the sign of
$\varepsilon$. It is then screened as $T$ is further lowered below
$T_{\mathrm{L}}$ when the system is governed by the infrared fixed point.
Notice that the moment formation region becomes smaller as $\left\vert
\varepsilon\right\vert $ is reduced and completely disappears at
$\varepsilon\simeq0$. In fact, over the whole mixed valence region
($\left\vert \varepsilon\right\vert \lesssim\Delta\equiv V^{2}/2$), valence
fluctuations suffice to prevent moment formation.

Another way of presenting the picture is to define two energy scales ($T_{s}$
and $T_{q}$) associated with the spin and quadrupolar degrees of freedom.
These scales cross each other in the intermediate valence region and
interchange roles as the high-temperature ($T_{\mathrm{H}}$) and
low-temperature ($T_{\mathrm{L}}$) scales that indicate the transition zones
among different regimes. The meeting of the two scales in the intermediate
valence region signals the absence of a moment formation regime between the
high and the low temperatures: the fixed point is reached \textit{without}
prior moment formation.

The rest of this article is organized as follows: in the next section we will
discuss the scattering matrices and the integrability of the model
(Sec.~\ref{S-matrix}); in the subsequent one we will present some necessary
formal developments in the theory of quantum inverse scattering and the
equations derived from them (Sec.~\ref{QISM}); in the following two we will
discuss the thermodynamics of the model --~we shall give formal derivations
(Sec.~\ref{TBA}) and analytic and numerical results (Sec.~\ref{ImpResults}%
)~--; in the last section we will provide a summary and a discussion of our
main results as well as an outlook of their applications in the theory of
heavy fermions (Sec.~\ref{END}).

\section{S-matrices and Integrability\label{S-matrix}}

The Hamiltonian $H_{\mathrm{MchA}}$ conserves the number of fermionic
excitations (electrons),
\[
N_{\mathrm{e}}=\sum_{\alpha,\sigma}\int\psi_{\alpha,\sigma}^{\dagger}\left(
x\right)  \psi_{\alpha,\sigma}\left(  x\right)  +\sum_{\sigma}f_{\sigma
}^{\dagger}f_{\sigma}%
\]
allowing us to study the system for an arbitrary but fixed value of
$N_{\mathrm{e}}$. We take $N_{\mathrm{e}}=0$ as our reference sector, an
$M$-degenerate eigenstate with energy $\varepsilon_{q}$. Our strategy will be
the usual one in coordinate Bethe-Ansatz: we solve the system for
$N_{\mathrm{e}}=1,2,\ldots$ and then generalize the solution to arbitrary
values of $N_{\mathrm{e}}$. Subsequently, as the $N_{\mathrm{e}}%
\rightarrow\infty$ limit is taken (discussed in later sections), the field
theory is recovered.

\subsection{Electron-Impurity Scattering Matrix}

When there is only one electron present in the system ($N_{\mathrm{e}}=1$),
the most general one-fermion state has the following form:
\[
\left\vert k\right\rangle =\sum_{\alpha,\beta,\sigma}\int F_{\alpha
\sigma;\beta}^{k}\left(  x\right)  \psi_{\alpha\sigma}^{\dagger}\left(
x\right)  \left\vert \bar{\beta}\right\rangle +G_{\sigma}^{k}\left\vert
\sigma\right\rangle ~\text{.}%
\]
To determine the eigenstates of the Hamiltonian in this sector, satisfying
$H\left\vert k\right\rangle =E^{k}\left\vert k\right\rangle $, we apply the
Hamiltonian to this generic state and derive the \textit{first quantized
Schr\"{o}dinger equations} in the sector. Using the expression $E^{k}%
=k+\varepsilon_{q}$ for the eigenenergies of eigenstates $\left\vert
k\right\rangle $, we read off the equations:
\[
\left\{
\begin{array}
[c]{c}%
\left(  -i\partial_{x}-k\right)  F_{\alpha\sigma;\beta}^{k}\left(  x\right)
+\delta\left(  x\right)  V\,\delta_{\alpha}^{\beta}G_{\sigma}^{k}=0\\
~\\
\left(  \varepsilon-k\right)  G_{\sigma}^{k}+\sum_{\alpha^{\prime}%
,\beta^{\prime}}V\,\delta_{\alpha^{\prime}}^{\beta^{\prime}}F_{\alpha^{\prime
}\sigma;\beta^{\prime}}^{k}\left(  0\right)  =0
\end{array}
\right.
\]
where the energy difference $\varepsilon=\varepsilon_{s}-\varepsilon_{q}$ was
introduced. We call $F_{\alpha\sigma;\beta}^{k}\left(  x\right)  $ and
$G_{\sigma}^{k}$ the wavefunctions. Eliminating $G_{\sigma}^{k}$ from the
first equation and setting $F_{\alpha\sigma;\beta}^{k}\left(  x\right)
=e^{ikx}\tilde{F}_{\alpha\sigma;\beta}^{k}\left(  x\right)  $ we have
(repeated indexes are summed over)
\[
\left(  k-\varepsilon\right)  \left(  -i\partial_{x}\right)  \tilde{F}%
_{\alpha\sigma;\beta}^{k}\left(  x\right)  +V^{2}\delta\left(  x\right)
\delta_{\alpha}^{\beta}\delta_{\alpha^{\prime}}^{\beta^{\prime}}\tilde
{F}_{\alpha^{\prime}\sigma;\beta^{\prime}}^{k}\left(  0\right)  =0~\text{.}%
\]
We make the following Bethe-type ansatz for the coordinate dependence of the
$\tilde{F}$-wavefunction: $\tilde{F}_{\alpha\sigma;\beta}^{k}\left(  x\right)
=\left[  \left(  \theta\left(  -x\right)  \mathbf{I}+\theta\left(  x\right)
\mathbf{S}\right)  A\right]  _{\alpha\sigma;\beta}^{k}$, where $\mathbf{S}$
will be called the electron-impurity scattering matrix and $A$ is an arbitrary
vector in the internal space of the one-electron sector. In order to fully
define the ansatz we adopt the following convention for the step function:
$\theta\left(  0\right)  =1/2$. Let us introduce the operator $\left[
\mathbf{Q}\right]  _{\alpha;\beta}^{\alpha^{\prime};\beta^{\prime}}%
=\delta_{\alpha}^{\beta}\delta_{\alpha^{\prime}}^{\beta^{\prime}}$ that acts
in flavor space and has the property $\mathbf{Q}^{2}=M\mathbf{Q}$ (recall $M$
is the number of values that the index $\alpha$ takes). Since $A$ is
arbitrary, one has a matrix equation for $\mathbf{S}$. Its solution is
\begin{align*}
\mathbf{S}_{1,0}  &  =\mathbf{I}_{1,0}\mathbf{-}\frac{i2V^{2}}{2\left(
k_{1}-\varepsilon\right)  +iMV^{2}}\mathbf{Q}_{1;0}\\
&  =\mathbf{I}_{1,0}+\frac{e^{-i\delta\left(  k_{1}-\varepsilon\right)  }%
-1}{M}\mathbf{Q}_{1;0}~\text{,}%
\end{align*}
where we used the index `$1$' for the only electron present in the system and
introduced the use of the index `$0$' for the impurity (the notation
$k_{0}\equiv\varepsilon$ will be also used latter). For the second way of
writing $\mathbf{S}_{1,0}$ we introduced the phase $\delta\left(
k-\varepsilon\right)  =2\arctan\frac{MV^{2}}{2(k-\varepsilon)}$.

It is easy to verify unitarity,
\[
\mathbf{SS}^{\dagger}=\mathbf{S}_{1,0}\mathbf{S}_{0,1}=\mathbf{I}~\text{.}%
\]

\subsection{Electron-Electron Scattering Matrix}

Consider now the case when there are two electrons present in the system
($N_{\mathrm{e}}=2$). The most general two-fermion state that we can write has
the following form (all indices are summed over):%
\begin{widetext}
\[
\left\vert k_{1}k_{2}\right\rangle =\int\int F_{\alpha_{1}\sigma_{1}\alpha
_{2}\sigma_{2};\beta}^{k_{1}k_{2}}\left(  x_{1},x_{2}\right)  \psi_{\alpha
_{1}\sigma_{1}}^{\dagger}\left(  x_{1}\right)  \psi_{\alpha_{2}\sigma_{2}%
}^{\dagger}\left(  x_{2}\right)  \left\vert \bar{\beta}\right\rangle +\int
G_{\alpha_{1}\sigma_{1};\sigma_{2}}^{k_{1}k_{2}}\left(  x_{1}\right)
\psi_{\alpha_{1}\sigma_{1}}^{\dagger}\left(  x_{1}\right)  \left\vert
\sigma_{2}\right\rangle ~\text{.}%
\]
Now again we apply the Hamiltonian in order to obtain the first quantized
Schr\"{o}dinger equations for eigenstates with eigenenergies $E^{k_{1}k_{2}%
}=k_{1}+k_{2}+\varepsilon_{q}$. We arrive at the following set of differential
equations:%
\[
\left\{
\begin{array}
[c]{c}%
\sum_{n=1,2}\left(  -i\partial_{x_{n}}-k_{n}\right)  \left[  \mathcal{A}%
F\right]  _{\alpha_{1}\sigma_{1}\alpha_{2}\sigma_{2};\beta}^{k_{1}k_{2}%
}\left(  x_{1},x_{2}\right)  +V\mathcal{A}\left[  \delta\left(  x_{2}\right)
\delta_{\alpha_{2}}^{\beta}G_{\alpha_{1}\sigma_{1};\sigma_{2}}^{k_{1}k_{2}%
}\left(  x_{1}\right)  \right]  =0\\
~\\
\left(  -i\partial_{x_{1}}-k_{1}+\varepsilon-k_{2}\right)  G_{\alpha_{1}%
\sigma_{1};0\sigma_{2}}^{k_{1}k_{2}}\left(  x_{1}\right)  +V\,\delta\left(
x_{2}\right)  \delta_{\alpha_{2}}^{\beta}\left[  \mathcal{A}F\right]
_{\alpha_{1}\sigma_{1}\alpha_{2}\sigma_{2};\beta}^{k_{1}k_{2}}\left(
x_{1},x_{2}\right)  =0
\end{array}
\right.
\]%
\end{widetext}%
where $\mathcal{A}=\mathbf{I}-\mathbf{P}^{xsq}=\mathbf{I}-\mathbf{P}%
^{x}\mathbf{P}^{s}\mathbf{P}^{q}$ is (twice) the antisymmetrizer in
coordinate, spin, and (quadrupolar)-flavor space, expressed in terms of
$\mathbf{P}^{x},\mathbf{P}^{s},\mathbf{P}^{q}$ the permutation operators that
act in the spaces indicated.

Now we make an ansatz for the $F$-wavefunction similar in spirit to the one we
made in the one-electron sector. There are six regions in configuration space,
corresponding to the six possible line arrangements of the impurity and the
two electrons. The region with, say, electron `$1$' to the immediate left of
the impurity is related via the S-matrix $\mathbf{S}_{1,0}$ to the region with
electron `$1$' to the immediate right (the position of electron `$2$' remains
unchanged). A similar role is played by $\mathbf{S}_{2,0}$. However, we still
need to determine how to relate the regions where electrons `$1$' and `$2$'
exchange places (away from the impurity).

Consider then,
\[
F_{\alpha_{1}\sigma_{1}\alpha_{2}\sigma_{2};\beta}^{k_{1}k_{2}}\left(
x_{1},x_{2}\right)  =e^{ik_{1}x_{1}+ik_{2}x_{2}}\tilde{F}_{\alpha_{1}%
\sigma_{1}\alpha_{2}\sigma_{2};\beta}^{k_{1}k_{2}}\left(  x_{1},x_{2}\right)
\]
with the expression for $\tilde{F}$ describing the six different arrangements
of particles,%
\begin{widetext}
\begin{align*}
\tilde{F}_{\alpha_{1}\sigma_{1}\alpha_{2}\sigma_{2};\beta}^{k_{1}k_{2}}\left(
x_{1},x_{2}\right)   &  =\left[  \left(  \theta\left(  x_{12}\right)
\theta\left(  -x_{2}\right)  \mathbf{I}+\theta\left(  x_{21}\right)
\theta\left(  -x_{1}\right)  \mathbf{S}+\theta\left(  -x_{1}\right)
\theta\left(  x_{2}\right)  \mathbf{S}_{2,0}+\right.  \right. \\
&  \left.  \left.  +\theta\left(  -x_{2}\right)  \theta\left(  x_{1}\right)
\mathbf{S}_{1,0}\mathbf{S}+\theta\left(  x_{1}\right)  \theta\left(
x_{12}\right)  \mathbf{S}_{1,0}\mathbf{S}_{2,0}+\theta\left(  x_{2}\right)
\theta\left(  x_{21}\right)  \mathbf{S}_{2,0}\mathbf{S}_{1,0}\mathbf{S}%
\right)  A\right]  _{\alpha_{1}\sigma_{1}\alpha_{2}\sigma_{2};\beta}%
^{k_{1}k_{2}}~\text{.}%
\end{align*}%
\end{widetext}%
Here $\mathbf{S}_{1,0}$ and $\mathbf{S}_{2,0}$ are the electron-impurity
S-matrices found above and $\mathbf{S\equiv S}_{1,2}$ is the electron-electron
S-matrix that we seek to determine in this subsection. As above $A$ is an
arbitrary vector in the internal space of the two-electron sector, determining
the state of two electrons and the impurity. To define unambiguously the
ansatz we adopt the regularization $\theta\left(  0^{-}\right)  \theta\left(
0^{+}\right)  \overset{\text{\textrm{\ae }}}{=}0$,\footnote{The equality is in
the sense of distributions (\textit{i.e.} \textit{almost everywhere}).}
consistent with the first order character of the differential equations. Let
us mention that this ansatz assumes that the same momenta $k_{1},k_{2}$
characterize the wavefunction in all six regions (\textit{i.e.}~orderings).
This is at the heart of the ansatz and will be shown to be valid later when we
discuss the Yang-Baxter conditions.

Inserting the wavefunction into the first of the Schr\"{o}dinger equations
above we verify after some algebra that the equation holds, determining
uniquely the form of the $G$-wavefunction. At this stage the electron-electron
scattering matrix remains arbitrary and we turn our attention to the second
Schr\"{o}dinger equation. Carrying out the algebra, we find that the equation
holds provided the following matrix constraint on the electron-electron
S-matrix is obeyed:
\[
\left(  \mathbf{S}_{2,0}-\mathbf{I}\right)  \left(  \mathbf{S}_{1,0}%
\mathbf{S}-\mathbf{I}\right)  -\mathbf{P}^{qs}\left(  \mathbf{S}%
_{1,0}-\mathbf{I}\right)  \left(  \mathbf{S}_{2,0}-\mathbf{S}\right)
=\mathbf{0}%
\]
where the matrices without indexes act on the internal space of the two
electrons. A careful examination of this equation reveals the presence of an
overall left-prefactor $\mathbf{Q}_{2;0}$. Since this operator is not
invertible, the solution of the constraint is not unique and there is still a
certain amount of freedom left in the choice of $\mathbf{S}$.

\subsubsection{An Integrable Solution}

The equation above does not have a unique solution for the two-electron
S-matrix $\mathbf{S}$. For our purpose we need to identify, however, a
particular solution having the appropriate physical properties that ensure the
generalization and consistency of the ansatz to any number of electrons
$N_{\mathrm{e}}$. These requirements are:%

\begin{table}[h] \centering
$%
\begin{tabular}
[c]{ccc}\hline\hline
Unitarity Condition & \qquad\qquad\qquad & $\mathbf{S}_{ij}\mathbf{S}%
_{ji}=\mathbf{I}$\\
Locality of the Scattering & \qquad\qquad\qquad & $\mathbf{S}_{ij}%
\mathbf{S}_{kl}=\mathbf{S}_{kl}\mathbf{S}_{ij}$\\
Yang-Baxter Relation & \qquad\qquad\qquad & $\mathbf{S}_{jk}\mathbf{S}%
_{ik}\mathbf{S}_{ij}=\mathbf{S}_{ij}\mathbf{S}_{ik}\mathbf{S}_{jk}%
$\\\hline\hline
\end{tabular}
\ $\caption{\label{YBE}Three necessary conditions for integrability.}%
\end{table}%

The first of these relations assures the reversibility of the scattering
processes (or reversibility of the scattering paths) and the other two are
enough to guarantee path independence for arbitrary $N_{\mathrm{e}}%
$.\cite{zjb66,yang67} With these conditions, any multiparticle scattering
process can be factorized into pairwise scattering events and there is no
ambiguity in the multiple ways of carrying out the factorization since they
are all equivalent (the reader can find an illustration of the situation in
four-particle space in the review article of Ref.~%
[\onlinecite{afl83}]%
). We remark that in the case of impurity models, any one of the indices in
these relations can take the value `$0$' that stands for the impurity.

These three conditions together with the constraint coming from the
Schr\"{o}dinger equation, constitute an over-constrained algebraic system for
determining the electron-electron S-matrix ($\mathbf{S\equiv S}_{1,2}$).
Nevertheless, it admits a solution.

It can be shown that the only solution is the following:
\[
\mathbf{S}_{1,2}=\frac{\left(  k_{1}-k_{2}\right)  -iV^{2}\mathbf{P}_{1,2}%
^{s}}{\left(  k_{1}-k_{2}\right)  -iV^{2}}\frac{\left(  k_{1}-k_{2}\right)
+iV^{2}\mathbf{P}_{1,2}^{q}}{\left(  k_{1}-k_{2}\right)  +iV^{2}}~\text{.}%
\]
This matrix serves to describe the electron band with linear dispersion within
a basis of reduced symmetry $SU\!\left(  NM\right)  \!\rightarrow SU\!\left(
N\right)  \otimes SU\!\left(  M\right)  $, consistent with the reduction in
symmetry operated by the addition of the impurity terms to the Hamiltonian of
the host band. It is important to emphasize that the introduction of
$\mathbf{S}_{1,2}$ does not signify that we have modified the original
Hamiltonian $H_{\mathrm{MchA}}$ by introducing electron-electron interaction.
Instead, the choice of $\mathbf{S}_{1,2}$ corresponds to a choice of basis in
the space of free electrons.\footnote{We remind the reader that the electron
gas with linear dispersion is highly degenerate. In other words, the
two-electron free Hamiltonian $h_{1,2}=-i(\partial_{1}+\partial_{2})$ has as a
general solution the wave function $\exp\left(  ik_{1}x_{1}+ik_{2}%
x_{2}\right)  [A_{a_{1}a_{2}}\theta(x_{1}-x_{2})+(\mathbf{S}_{1,2}%
A)_{a_{1}a_{2}}\theta(x_{2}-x_{1})]$, with $\mathbf{S}_{1,2}$ arbitrary. As
follows then from degenerate perturbation theory, one must choose already at
zeroth-order the correct combination with respect to which to turn on a given
perturbation. Our result corresponds to such a combination for the case when
the perturbation is a multi-channel Anderson impurity.}

\section{Periodic Boundary Conditions and Bethe-Ansatz equations\label{QISM}}

We proceed to impose boundary conditions. This is required in order to be able
to properly count and label the states. We will discuss here the case of
periodic boundary conditions, imposing the following set of conditions on the
$N_{\mathrm{e}}$-electron wave function,%
\[
\left.  F_{\left\{  \vec{m}\right\}  }^{\left\{  k\right\}  }\left(  \left\{
x\right\}  \right)  \right\vert _{x_{j}=\frac{L}{2}}=\left.  F_{\left\{
\vec{m}\right\}  }^{\left\{  k\right\}  }\left(  \left\{  x\right\}  \right)
\right\vert _{x_{j}=-\frac{L}{2}}%
\]
where we are considering a finite ring of length $L$. As we are able to move
electron $j$ to the far left ($x_{j}=-L/2$) or to the far right ($x_{j}=L/2$)
using the S-matrices, the boundary condition gives rise to the following
eigenvalue problem:%
\[
Z_{j}\vec{A}=z_{j}\vec{A}%
\]
where the eigenvalues $z_{j}=e^{-ik_{j}L}$ of the transfer matrix $Z_{j}$,
\[
Z_{j}=S_{jj-1}\ldots S_{j1}S_{j0}S_{jN}\ldots S_{jj+1}%
\]
will allow us to find the spectrum of the Hamiltonian via $E=\sum k_{j}$.
$\vec{A}$ is a vector in the internal space of the $N_{\mathrm{e}}$ electron
sector. Equivalently, this condition corresponds to taking particle $j$
\textit{around the ring} and asking that the wave function should not change;
it should hold for all $j=1,\ldots,N_{\mathrm{e}}$. The original problem of
finding the eigenenergies and eigenfunctions of the Hamiltonian is thus
reduced to that of finding the vector amplitudes in internal space ($\vec{A}$)
that are simultaneous eigenvectors of all these eigenvalue problems.

The solution of this class of eigenvalue problems was first tackled by
Yang,\cite{yang67} who solved the problem by means of a `second Bethe-Ansatz'
(see also the work of Baxter\cite{Baxter}). In the late 1970s the procedure
was systematized into what is known as the \textit{Quantum Inverse Scattering
Method} (QISM).\cite{ft79} However, the existing technology is insufficient
for our purpose; the structure of the impurity S-matrix in the multi-channel
Anderson model requires a reformulation and extension of the standard
formalism. We give a detailed account of those developments in one of the
appendices at the end of the article.

\subsection{Bethe-Ansatz Equations for the Multi-Channel Anderson Impurity
Model}

The eigenvalues, $z_{j}=e^{-ik_{j}L}$, of the transfer matrix are given in
terms of the \textit{charge rapidities} ($k_{j}$), which in their turn,
together with the \textit{spin rapidities} ($\Lambda_{\alpha}^{s\left(
r\right)  }$) and the \textit{quadrupolar-flavor rapidities} ($\Lambda
_{\alpha}^{q\left(  r\right)  }$), completely specify the particular
eigenstate (see first appendix). The spin rapidities ($\Lambda_{\alpha
}^{s\left(  r\right)  }$) and the {q-flavor rapidities} ($\Lambda_{\alpha
}^{q\left(  r\right)  }$) describe, respectively, the spin and flavor dynamics
as well as the symmetry of each state. The index $x=s$, ($q$) specifies that
$\Lambda_{\alpha}^{x\left(  r\right)  }$ refers to a spin (q-flavor) degree of
freedom, the index $r$ describes the \textit{rank} (related to the spin or
flavor symmetry of the state), and finally $\alpha$ labels the different
rapidities of each type and rank.

The charge, spin, and flavor rapidities must satisfy a set of equations -~the
Bethe-Ansatz Equations (BAE)~- that are derived in the first appendix. These
equations encode the full information of the model:
\[
e^{ik_{j}L}=\prod_{n=1}^{M_{1}^{s}}\frac{k_{j}-\Lambda_{n}^{s\left(  1\right)
}-i\Delta}{k_{j}-\Lambda_{n}^{s\left(  1\right)  }+i\Delta}\prod_{m=1}%
^{M_{1}^{q}}\frac{k_{j}-\Lambda_{m}^{q\left(  1\right)  }+i\Delta}%
{k_{j}-\Lambda_{m}^{q\left(  1\right)  }-i\Delta}%
\]
with the conditions,
\begin{align*}
\prod_{m\neq n}^{M_{r}^{s}}\frac{\Lambda_{n}^{s\left(  r\right)  }-\Lambda
_{m}^{s\left(  r\right)  }-i2\Delta}{\Lambda_{n}^{s\left(  r\right)  }%
-\Lambda_{m}^{s\left(  r\right)  }+i2\Delta}  &  =\prod_{\substack{m=1\\\sigma
=\pm1}}^{M_{r+\sigma}^{s}}\frac{\Lambda_{n}^{s\left(  r\right)  }-\Lambda
_{m}^{s\left(  r+\sigma\right)  }-i\Delta}{\Lambda_{n}^{s\left(  r\right)
}-\Lambda_{m}^{s\left(  r+\sigma\right)  }+i\Delta}\\
\prod_{m\neq n}^{M_{r}^{q}}\frac{\Lambda_{n}^{q\left(  r\right)  }-\Lambda
_{m}^{q\left(  r\right)  }-i2\Delta}{\Lambda_{n}^{q\left(  r\right)  }%
-\Lambda_{m}^{q\left(  r\right)  }+i2\Delta}  &  =\prod_{\substack{m=1\\\sigma
=\pm1}}^{M_{r+\sigma}^{q}}\frac{\Lambda_{n}^{q\left(  r\right)  }-\Lambda
_{m}^{q\left(  r+\sigma\right)  }-i\Delta}{\Lambda_{n}^{q\left(  r\right)
}-\Lambda_{m}^{q\left(  r+\sigma\right)  }+i\Delta}%
\end{align*}
where for convenience we have used the definitions,
\begin{align*}
\Lambda_{n}^{s,q\left(  0\right)  }  &  =k_{n}\\
\Lambda_{1}^{q\left(  M\right)  }  &  =\varepsilon
\end{align*}
and accordingly $M_{0}^{s,q}=N_{\mathrm{e}}$, $M_{M}^{q}=N_{\mathrm{i}}=1$,
and $M_{N}^{s}=0$. One sees that the effect of the impurity enters via the
auxiliary conditions for the \textit{q-flavor rapidities}. This is a
distinguishing feature, different from what happens in the equations for the
single-channel Anderson model, or in those for the Kondo model regardless of
the number of channels.

Each solution of the BAE corresponds to an eigenstate of the Hamiltonian. The
eigenfunction can in principle be determined from the eigenvector $\vec{A}$
and the corresponding energy eigenvalue is given by%
\[
E=\sum_{j}k_{j}~\text{.}%
\]
The charge-, spin- and quadrupolar-rapidities entering the solutions of the
BAE are in general complex, and take the form, in the thermodynamic limit, of
\textit{strings}.\cite{bethe31,takahashi72,ba99} An $n$-string of spin or
q-flavor rapidities consists of $n$ equally spaced complex numbers
symmetrically arranged around the real axis ($x=s,q$):
\[
\Lambda_{n\sigma}^{x\left(  r\right)  }=\Lambda_{n}^{x\left(  r\right)
}+i\left(  n+1-2\sigma\right)  \Delta\qquad\text{with}\quad\sigma
=1,\ldots,n~\text{.}%
\]
Strings are thus specified by a single real number $\Lambda_{n}^{x\left(
r\right)  }$, in terms of which we shall rewrite the BAE. Similarly, there
also complex values of the charge rapidities $k_{j}$ corresponding to bound
states among the bare electrons that build up the theory. They too form
strings and will be incorporated in the BAE through the real part
characterizing them. Consider a bound state of $n$ particles, the
corresponding charge rapidities form a string:
\[
k_{\sigma}^{\left(  0\right)  }=k^{\left(  n-1\right)  }+i\left(
n+1-2\sigma\right)  \Delta\qquad\text{with}\quad\sigma=1,\ldots,n
\]
and we have spin-rapidity strings associated with them in the first $n-1$
ranks: $k^{\left(  n-1\right)  }=\Lambda_{m}^{s\left(  r\right)  }$ for all
ranks such that $n=r+m$.

We now incorporate the string solutions into the equations and simplify them
using the notations and relations given in the second appendix. After some
algebra, one arrives at the final version of the discrete-real-BAE:%
\begin{widetext}%
\[
e^{ink_{j}^{\left(  n-1\right)  }L}=\prod_{m=1}\prod_{i=1}e_{nm}^{\prime
\prime}\left(  k_{j}^{\left(  n-1\right)  }-k_{i}^{\left(  m-1\right)
}\right)  \prod_{m=1}\prod_{\beta=1}e_{m}\left(  k_{j}^{\left(  n-1\right)
}-\Lambda_{m\beta}^{s\left(  n\right)  }\right)  \prod_{m=1}\prod_{\beta
=1}e_{nm}^{\prime\ast}\left(  k_{j}^{\left(  n-1\right)  }-\Lambda_{m\beta
}^{q\left(  1\right)  }\right)
\]
plus,
\begin{align*}
\prod_{m=1}\prod_{\beta=1}e_{nm}\left(  \Lambda_{n\alpha}^{s\left(  r\right)
}-\Lambda_{m\beta}^{s\left(  r\right)  }\right)   &  =\prod_{i=1}e_{n}\left(
\Lambda_{n\alpha}^{s\left(  r\right)  }-k_{i}^{\left(  r-1\right)  }\right)
\prod_{\sigma=\pm1}\prod_{m=1}\prod_{\beta=1}e_{nm}^{\prime}\left(
\Lambda_{n\alpha}^{s\left(  r\right)  }-\Lambda_{m\beta}^{s\left(
r+\sigma\right)  }\right) \\
\prod_{m=1}\prod_{\beta=1}e_{nm}\left(  \Lambda_{n\alpha}^{q\left(  1\right)
}-\Lambda_{m\beta}^{q\left(  1\right)  }\right)   &  =\prod_{m=1}\prod
_{i=1}e_{nm}^{\prime}\left(  \Lambda_{n\alpha}^{q\left(  1\right)  }%
-k_{i}^{\left(  m-1\right)  }\right)  \prod_{m=1}\prod_{\beta=1}e_{nm}%
^{\prime}\left(  \Lambda_{n\alpha}^{q\left(  1\right)  }-\Lambda_{m\beta
}^{q\left(  2\right)  }\right) \\
\prod_{m=1}\prod_{\beta=1}e_{nm}\left(  \Lambda_{n\alpha}^{q\left(  r\right)
}-\Lambda_{m\beta}^{q\left(  r\right)  }\right)   &  =\prod_{\sigma=\pm1}%
\prod_{m=1}\prod_{\beta=1}e_{nm}^{\prime}\left(  \Lambda_{n\alpha}^{q\left(
r\right)  }-\Lambda_{m\beta}^{q\left(  r+\sigma\right)  }\right)
\end{align*}%
\end{widetext}%
where we continue to use the notation $\Lambda_{1}^{q\left(  M\right)
}=\varepsilon$. The set of solutions of these equations determines the set of
eigenvalues of the problem. Now all the unknowns are real valued variables.

\subsubsection{Continuum Distributions Formulation}

One can study the system in the \textit{thermodynamic limit} where its size,
$L$, and the number of electrons, $N_{\mathrm{e}}$, both tend to infinity with
the density $N_{\mathrm{e}}/L$ being held fixed (one could later consider the
\textit{scaling limit}, when the density is allowed to go to infinity while
some physical scale is kept fixed).\cite{Andrei} In that case the separation
of distinct \textquotedblleft solutions\textquotedblright\ is of order
$O(1/N_{\mathrm{e}})$, so that rather than considering explicit
\textquotedblleft solutions\textquotedblright\ one can describe the system in
terms of \textquotedblleft densities of solutions\textquotedblright, typically
to be denoted by $\rho(z)$, describing the number of solutions falling in a
particular rapidity interval $(z,z+dz)$.\footnote{Notice that in order to
achieve a formulation in terms of continuum distributions, the essential limit
to be taken is $N_{\mathrm{e}}\rightarrow\infty$. We remark that this is the
same limit required in order to recover the field theory.}

To determine the densities one proceeds by taking logarithms of the BAE
obtaining transcendental equations for the rapidities characterized by
integers that arise from the logarithmic branches. These integers label the
\textquotedblleft solutions\textquotedblright\ and are the quantum numbers of
the eigenstate. One then constructs the \textit{counting functions}%
\cite{Andrei} for the different rapidities [in our case $\nu_{cn}\left(
z\right)  $, $\nu_{sn}^{\left(  r\right)  }\left(  z\right)  $, and $\nu
_{qn}^{\left(  r\right)  }\left(  z\right)  $]. The counting functions range
over all integers: those that have been selected for a state correspond to
\textquotedblleft solutions\textquotedblright, or \textquotedblleft
roots\textquotedblright, and those integers that are omitted correspond to
\textquotedblleft holes\textquotedblright. We denote by $\rho^{r}$ and
$\rho^{h}$ the various densities of roots and holes. For example $\rho
_{qn}^{r\left(  r\right)  }\left(  z\right)  $ denotes the density of roots of
rank-$r$ $n$-strings of quadrupolar rapidities. These density distributions
are related to the counting functions:
\begin{align*}
L^{-1}\partial_{z}\nu_{cn}\left(  z\right)   &  =\rho_{cn}\left(  z\right)
=\rho_{cn}^{r}\left(  z\right)  +\rho_{cn}^{h}\left(  z\right) \\
L^{-1}\partial_{z}\nu_{sn}^{\left(  r\right)  }\left(  z\right)   &
=\rho_{sn}^{\left(  r\right)  }\left(  z\right)  =\rho_{sn}^{r\left(
r\right)  }\left(  z\right)  +\rho_{sn}^{h\left(  r\right)  }\left(  z\right)
\\
L^{-1}\partial_{z}\nu_{qn}^{\left(  r\right)  }\left(  z\right)   &
=\rho_{qn}^{\left(  r\right)  }\left(  z\right)  =\rho_{qn}^{r\left(
r\right)  }\left(  z\right)  +\rho_{qn}^{h\left(  r\right)  }\left(  z\right)
\end{align*}
and should also obey the following relations,
\[
\left\{
\begin{array}
[c]{l}%
\frac{1}{L}N_{cn}=\int_{k}\rho_{cn}^{r}\left(  k\right) \\
~\\
\frac{1}{L}M_{sn}^{\left(  r\right)  }=\int_{\lambda}\rho_{sn}^{r\left(
r\right)  }\left(  \lambda\right) \\
~\\
\frac{1}{L}M_{qn}^{\left(  r\right)  }=\int_{\lambda}\rho_{qn}^{r\left(
r\right)  }\left(  \lambda\right)  ~\text{.}%
\end{array}
\right.
\]
These quantities are combined to define further ones,%
\[
\left\{
\begin{array}
[c]{l}%
N_{c}=\sum\limits_{n=1}n\,N_{cn}\\
~\\
M_{s}^{\left(  r\right)  }=\sum\limits_{n=1}n\,M_{sn}^{\left(  r\right)  }\\
~\\
M_{q}^{\left(  r\right)  }=\sum\limits_{n=1}n\,M_{qn}^{\left(  r\right)  }%
\end{array}
\right.
\]
where $N_{c}\equiv N_{\text{\textrm{e}}}$ corresponds to the number of
electrons in the system. Further, the energy (that coincides with the momentum
for a system with linear dispersion) and the number of particles with
different spins or q-flavors are given by
\[
\left\{
\begin{array}
[c]{l}%
\frac{1}{L}E=\frac{1}{L}\varepsilon_{q}+\sum\limits_{n=1}n\int_{k}k\,\rho
_{cn}^{r}\left(  k\right) \\
~\\
m_{sr}=M_{s}^{\left(  r-1\right)  }-M_{s}^{\left(  r\right)  }+\sum
\limits_{m=r}N_{cm}\\
~\\
m_{qr}=M_{q}^{\left(  r-1\right)  }-M_{q}^{\left(  r\right)  }+\delta
_{r,1}N_{c}%
\end{array}
\right.
\]
where $r=1,\ldots,X$ for $SU\!\left(  X\right)  $ ($X=M,N$) and we remind the
reader that $M_{q}^{\left(  M\right)  }=N_{\text{\textrm{i}}}=1$ is the number
of impurities in the system. Consistently with this convention we have the
density $\rho_{q1}^{r\left(  M\right)  }\left(  \lambda\right)  =L^{-1}%
\delta\left(  \lambda-\varepsilon\right)  $. We will use these quantities to
couple to crystal fields when we compute the thermodynamics.

Starting from the derivatives of the logarithm of the BAE and using the
density distributions defined just above plus the convolution kernels defined
in the second appendix one can write the continuum version of the BAE:%
\begin{align*}
\rho_{cn}^{h}=  &  \frac{n}{2\pi}-C_{nm}\ast\rho_{cm}^{r}-K_{m}\ast\rho
_{sm}^{r\left(  n\right)  }+B_{nm}\ast\rho_{qm}^{r\left(  1\right)  }\\
\rho_{sn}^{h\left(  r\right)  }=  &  -A_{nm}\ast\rho_{sm}^{r\left(  r\right)
}+K_{n}\ast\rho_{cr}^{r}+\\
&  \qquad\qquad\qquad+B_{nm}\ast\rho_{sm}^{r\left(  r-1\right)  }+B_{nm}%
\ast\rho_{sm}^{r\left(  r+1\right)  }\\
\rho_{qn}^{h\left(  r\right)  }=  &  -A_{nm}\ast\rho_{qm}^{r\left(  r\right)
}+\delta_{r,1}B_{nm}\ast\rho_{cm}^{r}+\\
&  \qquad\qquad\qquad+B_{nm}\ast\rho_{qm}^{r\left(  r-1\right)  }+B_{nm}%
\ast\rho_{qm}^{r\left(  r+1\right)  }%
\end{align*}
(repeated indices are contracted). The way they are written, these equations
determine the densities of holes as a function of the densities of roots and
their solutions correspond to the different eigenvalues of the system.

A detailed analysis of these equations will be presented elsewhere. In the
following we will use them as a starting point to write a second set of
equations whose solution allows one to compute the free energy of the system
and gives that way access to all thermodynamic quantities.

\section{Thermodynamic Bethe-Ansatz\label{TBA}}

The extension of the Bethe-Ansatz formalism to obtain finite temperature
information was first done in the case of the Bose gas by C.~N.~and
C.~P.~Yang.\cite{yy69} It was later adapted to the study of spin chains in the
works of Gaudin,\cite{gaudin71,Gaudin} Takahashi,\cite{takahashi71,Takahashi}
and others. The formalism is by now well developed and goes under the name of
Thermodynamic Bethe-Ansatz (TBA). In the context of impurity models it was
extensively used to find the impurity contributions to different thermodynamic
quantities. In the following we outline the main steps and results of the TBA
procedure as it applies to the multi-channel Anderson impurity model.

We shall proceed in a standard manner and derive an expression for the free
energy of the system. We will work in the \textit{grand canonical}
\textit{ensemble}, where the free energy is defined as\footnote{See Ref.~%
[\onlinecite{Andrei}]
for a proof that it is a good approximation to restrict ourselves to the
s-channel or q-channel highest weight states in the case of $SU\left(
2\right)  $; this proof can be extended to the general $SU\left(  N\right)  $
case.}
\[
F=E-TS-\mu_{c}N_{c}-\sum_{r}h_{sr}m_{sr}-\sum_{r}h_{qr}m_{qr}~\text{.}%
\]
Out of all the elements that enter this expression, the only one not mentioned
in the previous section, and requiring special attention here, is the entropy.
We define the entropy of a density distribution as follows (subindices are
suppressed):
\[
\frac{1}{L}\mathcal{S}\left\{  \rho^{r},\rho^{h}\right\}  =\rho^{r}\ln\left(
1+\eta\right)  +\rho^{h}\ln\left(  1+\bar{\eta}\right)
\]
where we introduced the notational conventions $\eta=\rho^{h}/\rho^{r}$ and
$\bar{\eta}=\eta^{-1}=\rho^{r}/\rho^{h}$. $\mathcal{S}\left\{  \rho^{r}%
,\rho^{h}\right\}  $ measures the number of microscopic states consistent with
a given macroscopic (thermodynamic) state given by $\rho^{r}$ and $\rho^{h}%
$.\cite{yy69} For the sake of further compactness, we also introduce the
notations $\mathrm{f}=\ln\left(  1+\eta\right)  $ and $\mathrm{\bar{f}}%
=\ln\left(  1+\bar{\eta}\right)  $. We write down the explicit expression of a
\textit{free energy functional} for the system:%
\begin{widetext}%
\begin{align*}
\frac{1}{L}\mathcal{F}\left\{  \rho_{x}^{r,h}\right\}  =\frac{1}{L}%
\varepsilon_{q}+\frac{1}{L}h_{qM}  &  +\sum_{n}\int_{k}\Bigl[n\left(
k-\mu_{c}-h_{q1}\right)  -\sum_{i\leq n}h_{si}-T~\mathrm{f}_{cn}%
\Bigr]\rho_{cn}^{r}-\sum_{n}\int_{k}T~\mathrm{\bar{f}}_{cn}\rho_{cn}^{h}+\\
&  +\sum_{r,n}\int_{\lambda}\left[  -\left(  h_{sr+1}-h_{sr}\right)
n-T~\mathrm{f}_{sn}^{\left(  r\right)  }\right]  \rho_{sn}^{r\left(  r\right)
}-\sum_{r,n}\int_{\lambda}T~\mathrm{\bar{f}}_{sn}^{\left(  r\right)  }%
\rho_{sn}^{h\left(  r\right)  }+\\
&  +\sum_{r,n}\int_{\lambda}\left[  -\left(  h_{qr+1}-h_{qr}\right)
n-T~\mathrm{f}_{qn}^{\left(  r\right)  }\right]  \rho_{qn}^{r\left(  r\right)
}-\sum_{r,n}\int_{\lambda}T~\mathrm{\bar{f}}_{qn}^{\left(  r\right)  }%
\rho_{qn}^{h\left(  r\right)  }~\text{.}%
\end{align*}%
\end{widetext}%
We next seek to determine the free energy, that in the thermodynamic limit we
are allowed to evaluate as a saddle point: we vary $\mathcal{F}$ with respect
to the distributions $\rho^{r}$ and $\rho^{h}$, subject to the constraint that
they must satisfy the BAE. We thus obtain the so-called Thermodynamic
Bethe-Ansatz equations:
\begin{align*}
\mathrm{f}_{cn}=  &  n\left(  k-\mu_{c}-h_{q1}\right)  /T-\delta_{n\geq
i}h_{si}/T+\\
&  \qquad\qquad+C_{nm}\ast\mathrm{\bar{f}}_{cm}-K_{m}\ast\mathrm{\bar{f}}%
_{sm}^{\left(  n\right)  }-B_{nm}\ast\mathrm{\bar{f}}_{qm}^{\left(  1\right)
}\\
\mathrm{f}_{sn}^{\left(  r\right)  }=  &  n\left(  h_{sr}-h_{sr+1}\right)
/T+K_{n}\ast\mathrm{\bar{f}}_{cr}+A_{nm}^{rs}\ast\mathrm{\bar{f}}%
_{sm}^{\left(  s\right)  }\\
\mathrm{f}_{qn}^{\left(  r\right)  }=  &  n\left(  h_{qr}-h_{qr+1}\right)
/T-\delta_{1}^{r}B_{nm}\ast\mathrm{\bar{f}}_{cm}+A_{nm}^{rs}\ast
\mathrm{\bar{f}}_{qm}^{\left(  s\right)  }%
\end{align*}
(repeated indices are contracted and the kernels $A_{nm}^{rs}$ are given in
the second appendix). It is possible to reformulate the TBA equations as a set
of recursions linking the different unknown distributions.\cite{Takahashi}
Such a formulation does not involve infinite sums and has also the virtue of
rendering the structure of the problem more transparent. After a few algebraic
manipulations --~making extensive use of the recursion relations for the
convolution kernels mentioned in the second appendix~-- one can re-express the
TBA equations in what we call their \textit{recursive formulation} (sometimes
referred to as the Gaudin-Takahashi form). They read:\footnote{We introduce
the notation $\hat{\delta}_{n,1}$ that stands for $\left(  1-\delta
_{n,1}\right)  $.}%
\begin{widetext}%
\[
\left\{
\begin{array}
[c]{l}%
\qquad\text{Equations for the spin-rapidities:}\\
~\\
\qquad\ \ \mathrm{f}_{sn}^{\left(  r\right)  }=\operatorname{lex}\left(
\delta_{n,1}G\ast\mathrm{\bar{f}}_{cr}+G\ast\mathrm{f}_{sn+1}^{\left(
r\right)  }+\hat{\delta}_{n,1}G\ast\mathrm{f}_{sn-1}^{\left(  r\right)
}-G\ast\mathrm{\bar{f}}_{sn}^{\left(  r+1\right)  }-G\ast\mathrm{\bar{f}}%
_{sn}^{\left(  r-1\right)  }\right) \\
~\\
\qquad\text{Equations for the flavor-rapidities:}\\
~\\
\qquad\ \ \mathrm{f}_{qn}^{\left(  r\right)  }=\operatorname{lex}\left(
-\delta_{n\leq N}\delta_{r,1}G\ast\mathrm{\bar{f}}_{cn}+G\ast\mathrm{f}%
_{qn+1}^{\left(  r\right)  }+\hat{\delta}_{n,1}G\ast\mathrm{f}_{qn-1}^{\left(
r\right)  }-G\ast\mathrm{\bar{f}}_{qn}^{\left(  r+1\right)  }-G\ast
\mathrm{\bar{f}}_{qn}^{\left(  r-1\right)  }\right) \\
~\\
\qquad\text{Equations for the charge-rapidities:}\\
~\\
\qquad\left\{
\begin{array}
[c]{l}%
\qquad\mathrm{\bar{f}}_{cn_{<N}}=\operatorname{lex}\left(  G\ast
\mathrm{\bar{f}}_{qn}^{\left(  1\right)  }+G\ast\mathrm{f}_{s1}^{\left(
n\right)  }-G\ast\mathrm{f}_{cn+1}-\hat{\delta}_{n,1}G\ast\mathrm{f}%
_{cn-1}\right) \\
~\\
\qquad\text{plus the `driving' equation,}\\
~\\
\qquad\left\{
\begin{array}
[c]{c}%
\mathrm{f}_{cN}=\operatorname{lex}\left(  N\frac{\lambda}{\tau}-{}^{R}%
G_{t}^{\left(  M,M\right)  }\ast\mathrm{f}_{qN}^{\left(  M-t\right)  }+{}%
^{R}G_{m}^{\left(  N,M\right)  }\ast\mathrm{\bar{f}}_{cm}\right) \\
~\\
\ \ \ \mathrm{\bar{f}}_{cN}=\operatorname{lex}\left(  -M\frac{\lambda}{\tau
}-{}^{R}G_{m_{<N}}^{\left(  N,N\right)  }\ast\mathrm{\bar{f}}_{cm}+{}%
^{R}G_{t_{\leq M}}^{\left(  M,N\right)  }\ast\mathrm{f}_{qN}^{\left(
M-t\right)  }\right)
\end{array}
\right.
\end{array}
\right.
\end{array}
\right.
\]%
\end{widetext}%
\begin{figure*}[htb]
\begin{center}%
\raisebox{-0cm}{\includegraphics[
height=11.8047cm,
width=15.0293cm
]%
{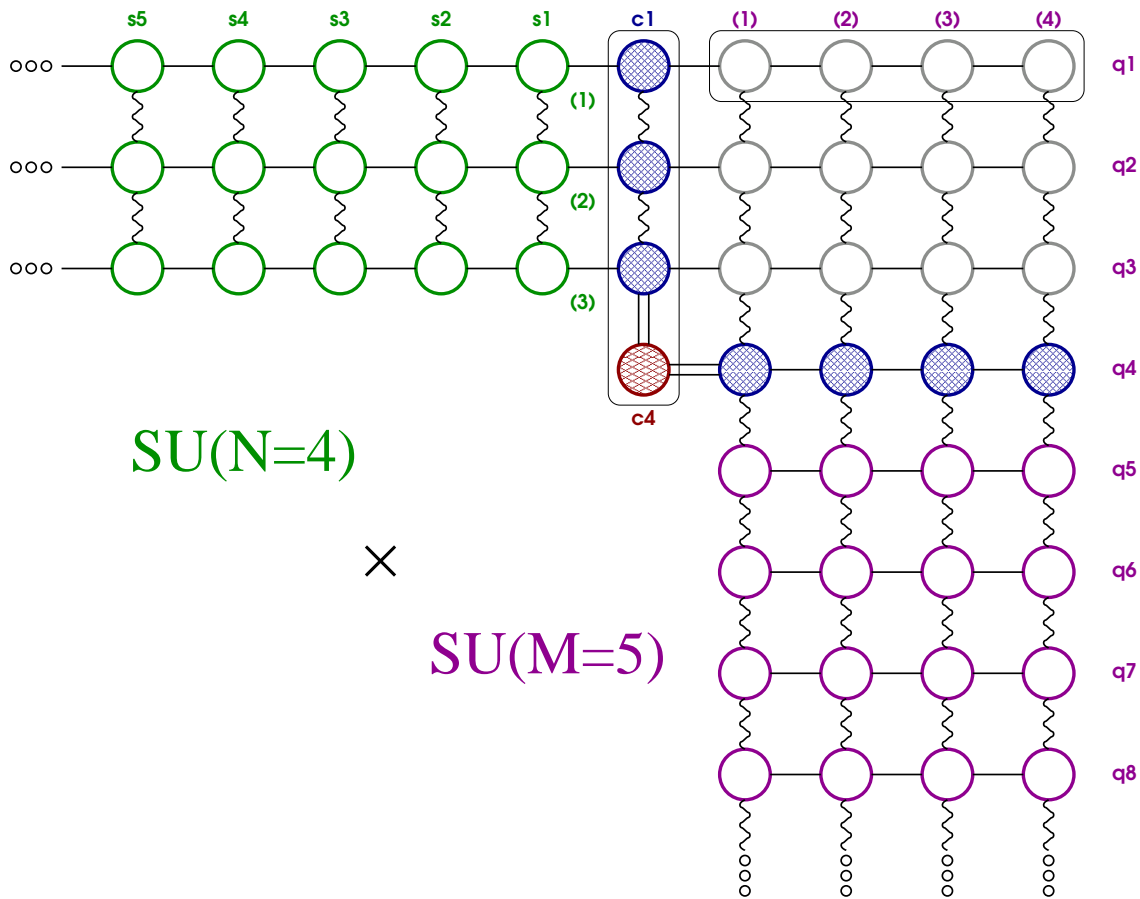}%
}%
\caption{\label{mchtbas}TBA diagram for a particular realization of the multi-channel
Anderson model in which the symmetry of the impurity is
$SU\!\left(N\right) \otimes SU\!\left(M\right)$ with $N=4$ and $M=5$.
A detailed explanation of this diagram is given in the text.}%
\end{center}
\end{figure*}%
where the dimensionless variables $\lambda=\pi(k-\mu)/2\Delta$ and $\tau=\pi
T/4\Delta$ were introduced\footnote{Here $\mu$ is the total chemical potential
acting on the electrons, defined as $\mu=\mu_{c}+\mu_{s}+\mu_{q}$ with
$\mu_{s}=\frac{1}{N}\sum_{n=1}^{N}h_{sn}$ and $\mu_{q}=\frac{1}{M}\sum
_{m=1}^{M}h_{qm}$.} and, for convenience, we defined the function
$\operatorname{lex}\left(  x\right)  =\ln\left(  1+\exp\left(  x\right)
\right)  $. The general recursion kernels ($G_{m}^{\left(  N,M\right)  }$) and
their regularized versions (${}^{R}G_{m}^{\left(  N,M\right)  }$) are
discussed in the second appendix. The expressions for $\mathrm{f}_{cN}$ and
$\mathrm{\bar{f}}_{cN}$ can of course be derived from each other, but here we
write down both of them explicitly because both are useful in the numerical studies.

These equations are complemented by asymptotic conditions that are also a
direct consequence of the `non-recursive' TBA equations:%
\[
K_{n+1}\ast\mathrm{f}_{xn}^{\left(  r\right)  }-K_{n}\ast\mathrm{f}%
_{xn+1}^{\left(  r\right)  }\underset{n\rightarrow\infty}{\longrightarrow
}\alpha_{x}^{r}\equiv\frac{h_{xr+1}-h_{xr}}{T}%
\]
where $h_{xr}$ is either $h_{sr=1,\ldots,N}$ or $h_{qr=1,\ldots,M}$ for the
spin or quadrupolar $\mathrm{f}$-distributions, respectively.

The structure of the recursive relations between the different distributions
can be nicely visualized in a graph. In Fig.~\ref{mchtbas} we show an example
of such a graph, a `TBA-diagram' for a particular realization of the
multi-channel Anderson model.

This representation of the equations highlights their connectivity and is the
natural extension of the graphical representation commonly used in the case of
other integrable models (for comparison and an example of another related
impurity model see the work on the multi-channel Coqblin-Schrieffer
model\cite{jaz98}). The different nodes correspond to the different
distributions: (i) those in the horizontal stripe extending towards the left
represent the $\eta_{sn}^{\left(  r\right)  }$, (ii) those in the vertical
stripe extending downwards represent the $\bar{\eta}_{qn}^{\left(  r\right)
}$, and (iii) those in the vertical column enclosed in a box correspond to the
$\bar{\eta}_{cn}$. The graph was drawn for the particular case of $SU\!\left(
N=4\right)  \otimes SU\!\left(  M=5\right)  $ symmetry, but its structure is
generic. The horizontal straight lines indicate that the nodes are two-way
connected by the equations according to $\ln\eta_{\mathrm{node}}=G\ast
T\ln\left(  1+\eta_{\mathrm{neighbor}}\right)  $ and the vertical wavy lines
indicate that the two-way connections are given by $\ln\eta_{\mathrm{node}%
}=-G\ast T\ln\left(  1+\bar{\eta}_{\mathrm{neighbor}}\right)  $. The double
straight lines highlight the special kind of connections in the case of
$\bar{\eta}_{cN}$.

There is a dual interpretation of this graph in terms of the reciprocals of
all the distributions. Within this new picture the different nodes are as
follows: (i) those in the horizontal stripe extending towards the left
represent the $\bar{\eta}_{sn}^{\left(  r\right)  }$, (ii) those in the
vertical stripe extending downwards represent the $\eta_{qn}^{\left(
r\right)  }$, and (iii) those in the vertical column enclosed in a box
correspond to the $\eta_{cn}$. The meaning of the straight and wavy lines also
gets interchanged: horizontal straight lines now mean $\ln\eta_{\mathrm{node}%
}=-G\ast T\ln\left(  1+\bar{\eta}_{\mathrm{neighbor}}\right)  $ and vertical
wavy lines mean $\ln\eta_{\mathrm{node}}=G\ast T\ln\left(  1+\eta
_{\mathrm{neighbor}}\right)  $. This duality plays a role when one considers
the two different integral valence limits, such limits will be discussed later.

The two boxes shown in the graph enclose those nodes that enter the expression
for the impurity contribution to the free energy. Starting from the free
energy functional and evaluating it using the distributions that obey the TBA
equations, after a certain amount of algebra, one arrives at the expression
$F=F_{\mathrm{bulk}}+F_{\mathrm{imp}}$ where%
\[
F_{\mathrm{bulk}}=-\frac{L}{2\pi}\sum_{n}n\int_{k}T\ln\left(  1+\bar{\eta
}_{cn}\right)
\]
and%
\begin{multline*}
F_{\mathrm{imp}}=\varepsilon_{q}+\mu_{q}-\left[  \sum_{t}G_{t}^{\left(
M,M\right)  }\ast T\ln\left(  1+\eta_{q1}^{\left(  t\right)  }\right)
\right.  +\\
+\left.  \sum_{m}G_{1}^{\left(  M+m,M\right)  }\ast T\ln\left(  1+\bar{\eta
}_{cm}\right)  \right]  _{k=\varepsilon}%
\end{multline*}
from which all the different thermodynamic quantities of interest can be
derived. Here $F_{\mathrm{bulk}}$ is the bulk contribution to the free energy
(an extensive part that recovers the standard result for free electrons and is
all there is in the absence of the impurity), on the other hand
$F_{\mathrm{imp}}$ is the extra contribution due to the presence of the
impurity. In the following we shall pursue the study of the latter.

As expected for an infinite flat-band, the bulk part of the free energy is
found to be divergent. To study it, some form of regularization should be
introduced, \textit{e.g.}~a bandwidth cut-off. A convenient procedure that was
successfully applied in the Bethe-Ansatz study of the single-channel
degenerate Anderson model is the introduction of a `Lorentzian cut-off
scheme'.\cite{schlottmann83a,schlottmann83b,ca86} In that framework it is
usual to fix the density and let the chemical potential be determined as an
implicit function of the cut-off parameter. As a result the impurity
thermodynamics is found to be naturally expressible in terms of a certain
\textit{scaling invariant} parameter, $\varepsilon^{\ast}\left(
\varepsilon,\Delta\right)  $,\cite{haldane78} an implicit function of cut-off,
temperature, and fields.\cite{ca86} On the other hand, the impurity
contribution to the free energy is found to be regular (\textit{i.e.}~finite
even in the limit of infinite cut-off). Since we shall be interested solely in
the impurity thermodynamics, we will introduce no cut-off and continue to work
always in the grand canonical ensemble, keeping fixed the chemical potential
rather than the density. With this convention, we can continue to describe the
physics in terms of the original microscopic parameter $\varepsilon$.

\section{Impurity Thermodynamics\label{ImpResults}}

In this section we study the impurity contributions to the different
thermodynamic quantities of interest. We will first list some analytical
results and then give an extended discussion of the numerical solution of the
TBA equations and the results obtained for several thermodynamic quantities
across the different regimes of the model.

\subsection{Analytical Results}

As already pointed out in the two-channel case,\cite{ba02} the multi-channel
Anderson model displays a non-trivial zero temperature limit for the impurity
contribution to the system entropy. This is a clear indication of the
non-Fermi liquid character of the ground state. We can find the value of this
entropy and identify the relevant scale for the cross-over into the low
temperature phase in closed analytical form and for the general multi-channel case.

\subsubsection{Zero Temperature Solution\label{Tzero}}

We begin the study of the TBA equations by taking the zero temperature limit
of the equations in order to identify the ground state. This is a required
preliminary step before abording the study of the impurity contribution to the
residual entropy.

It is convenient to introduce the distributions $\xi\equiv T\ln\eta$. Assuming
that the derivatives $\partial_{T}\xi$ are bounded distributions, or have at
most isolated logarithmic divergencies, we derive the following
limits:\cite{Takahashi}%
\begin{align*}
\lim_{T\rightarrow0^{+}}T\ln\left(  1+\eta\right)   &  =\left\vert
\xi\right\vert \;\theta\left(  \xi\right)  \equiv\xi^{+}\\
\lim_{T\rightarrow0^{+}}T\ln\left(  1+\bar{\eta}\right)   &  =\left\vert
\bar{\xi}\right\vert \;\theta\left(  \bar{\xi}\right)  \equiv\bar{\xi}%
^{+}=-\xi^{-}~\text{.}%
\end{align*}
These limits are the key step in the derivation; using them it is immediate to
write down the zero temperature limit of the equations,
\begin{align*}
NMz=  &  C_{Mt_{\leq M}}\ast\xi_{qN}^{\left(  M-t\right)  +}-C_{Nm}\ast
\bar{\xi}_{cm}^{+}\\
\bar{\xi}_{cn_{<N}}=  &  G\ast\left[  \bar{\xi}_{qn}^{\left(  1\right)  +}%
+\xi_{s1}^{\left(  n\right)  +}-\xi_{cn+1}^{+}-\hat{\delta}_{n,1}\xi
_{cn-1}^{+}\right] \\
\xi_{sn}^{\left(  r\right)  }=  &  \delta_{n,1}G\ast\bar{\xi}_{cr}^{+}+\\
&  +G\ast\left[  \xi_{sn+1}^{\left(  r\right)  +}+\hat{\delta}_{n,1}\xi
_{sn-1}^{\left(  r\right)  +}-\bar{\xi}_{sn}^{\left(  r+1\right)  +}-\bar{\xi
}_{sn}^{\left(  r-1\right)  +}\right] \\
\xi_{qn}^{\left(  r\right)  }=  &  -\delta_{n\leq N}\,\delta_{r,1}G\ast
\bar{\xi}_{cn}^{+}+\\
&  +G\ast\left[  \xi_{qn+1}^{\left(  r\right)  +}+\hat{\delta}_{n,1}\xi
_{qn-1}^{\left(  r\right)  +}-\bar{\xi}_{qn}^{\left(  r+1\right)  +}-\bar{\xi
}_{qn}^{\left(  r-1\right)  +}\right]
\end{align*}
where the variable is the shifted rapidity $z=k-\mu$, and we use the notation
$\hat{\delta}_{n,1}=1-\delta_{n,1}$. These equations can be solved exactly
when $M=N$ and there are no external applied fields. Let us start discussing
the two-channel case ($M=N=2$). Since there is only one rank we drop the rank
superindex during this discussion. It is easy to see from the equations that
$\xi_{sn}^{-}=0$ $\forall n$ and $\xi_{qn}^{-}=0$ $\forall n>2$. It is natural
to assume in the zero fields case that $\xi_{c1}^{-}=0$, which implies
$\xi_{q1}^{-}=0$. Notice that after these considerations the equations for the
spin rapidities are decoupled from the rest. We solve them first, taking into
account the asymptotic condition $K_{n+1}\ast\xi_{sn}-K_{n}\ast\xi
_{sn+1}\rightarrow-2h_{s}=0$ (we use the standard definition for the magnetic
field). The solution is very simple: put all $\xi_{sn}$'s to zero. We see also
that in analogy to the spin case we can solve for the q-flavor sector simply
by taking all $\xi_{qn}^{+}$'s with $n\neq2$ to be zero. Only three non-zero
distributions remain and are given directly by the equations%
\[
\left\{
\begin{array}
[c]{c}%
\xi_{c2}=\xi_{c2}^{+}+\xi_{c2}^{-}=2z\\
~\\
\left\{
\begin{array}
[c]{c}%
\xi_{c1}=\xi_{c1}^{+}=G\ast\xi_{c2}^{+}\\
~\\
\xi_{q2}=\xi_{q2}^{-}=G\ast\xi_{c2}^{-}~\text{.}%
\end{array}
\right.
\end{array}
\right.
\]
Thus, in the absence of external fields, the ground state is built out of a
sea of charge-spin strings filled up to the chemical potential (\textit{i.e.}%
~$k=\mu$) and a completely filled sea of q-flavor 2-strings. We can derive the
zero temperature impurity valence, or level occupancy, in a closed form (we
use the notations $\lambda=\pi z/2\Delta$ and $J=2\Delta/\left(
\varepsilon-\mu\right)  $, notice the charge susceptibility can be easily
obtained taking a derivative):
\[
n_{c,\mathrm{imp}}^{0}=\int_{-\infty}^{+\infty}\frac{4\left(  \lambda
-\pi/J\right)  }{\left[  \left(  \lambda-\pi/J\right)  ^{2}+\pi^{2}\right]
^{2}}~\xi_{q2}\left(  \lambda\right)  ~d\lambda~\text{.}%
\]
The occupancy is integral (\textit{i.e.} $n_{c}^{0}\approx0,1$) for
$\left\vert \varepsilon-\mu\right\vert \gg\Delta$, and non-integral otherwise.
It is a simple exercise to show that the impurity contribution to the residual
entropy is $S_{\mathrm{imp}}^{0}=\ln\sqrt{2}$ --~a result that is consistent
with what was found in the integral valence limit given by the two-channel
Kondo model.\cite{ad84} We will give below a derivation of $S_{\mathrm{imp}%
}^{0}$ for the general multi-channel case.

Let us now return to the case with arbitrary values of $M$ and $N$, but always
in the absence of applied external fields. The equations are more involved and
one is not able to make general statements about the positivity of the
different distributions the way we did in the two-channel case. We proceed by
making the following educated guess:\medskip

\begin{enumerate}
\item $\xi_{sn}^{\left(  r\right)  }=0\quad\forall n,r$ (this is to ensure a
paramagnetic ground state), and it implies $\bar{\xi}_{cr_{<N}}^{+}=0$.

\item $\xi_{qn}^{\left(  r\right)  }=0\quad\forall n\neq N$ and for $n=N$ the
distributions are negative, \textit{i.e.} $\xi_{qN}^{\left(  r\right)  +}%
=0$.\medskip
\end{enumerate}

The equations then take the form,
\begin{align*}
NMz  &  =C_{MM}\ast\xi_{cN}^{+}-C_{NN}\ast\bar{\xi}_{cN}^{+}\\
\xi_{cn_{<N}}^{+}  &  =G\ast\left[  \xi_{cn+1}^{+}+\hat{\delta}_{n,1}%
\xi_{cn-1}^{+}\right] \\
\bar{\xi}_{qN}^{\left(  r\right)  +}  &  =\delta_{r,1}G\ast\bar{\xi}_{cN}%
^{+}+G\ast\left[  \bar{\xi}_{qN}^{\left(  r+1\right)  +}+\bar{\xi}%
_{qN}^{\left(  r-1\right)  +}\right]  ~\text{.}%
\end{align*}
The first equation is decoupled from the rest and one should solve it to
determine $\xi_{cN}$ that will later play the role of a driving term on the
other equations. A solution that is exact when $N=M$, but approximate
otherwise, is given by $\xi_{cN}=z\,\left[  M\,\theta\left(  -z\right)
+N\,\theta\left(  z\right)  \right]  $. When approximate, this solution is not
accurate in the rapidity interval $\left\vert z\right\vert <\Delta$. This
region corresponds, in the free energy, to the mixed valence regime that
interpolates between two different multi-channel Coqblin-Schrieffer limits
(this will be discussed more in detail in Secs.~\ref{SW} and \ref{END}). When
$N\neq M$, the above solution can be used as the initial guess for an
iterative numerical scheme; this idea will also be useful for the numerical
solution of the finite temperature case. The remaining two sets of equations
can then be `unnested' to obtain explicit expressions for the different
remaining distributions, all of which are fully determined by $\xi_{cN}$ (in
complete analogy to the two-channel case).

\subsubsection{Residual Impurity Entropy\label{ResEntropy}}

Regardless of its precise form, the solution for $\xi_{cN}$ will have both
positive and negative non-zero parts (as is clear from considering the large
$\left\vert z\right\vert $ asymptotic limits). Thus, the zero temperature and
fields solution discussed earlier indicates that $\xi_{cn_{<N}}^{+}$ and
$\bar{\xi}_{qN}^{\left(  r\right)  +}$ are non-zero everywhere.\footnote{The
notation in the subindex means: an arbitrary $n$ such that $n<N$.} In turn
this means that $\bar{\eta}_{cn_{<N}}$ and $\eta_{qN}^{\left(  r\right)  }$
are zero in the zero temperature limit. At the level of the TBA equations,
this has the effect of isolating the equations for the $\eta_{qn_{<N}%
}^{\left(  r\right)  }$ distributions:
\begin{align*}
\ln\eta_{qn_{<N}}^{\left(  r\right)  }=G\ast &  \left[  \hat{\delta}%
_{n,N-1}\ln\left(  1+\eta_{qn+1}^{\left(  r\right)  }\right)  +\right. \\
&  +\hat{\delta}_{n,1}\ln\left(  1+\eta_{qn-1}^{\left(  r\right)  }\right)
-\\
&  \left.  -\ln\left(  1+\bar{\eta}_{qn}^{\left(  r+1\right)  }\right)
-\ln\left(  1+\bar{\eta}_{qn}^{\left(  r-1\right)  }\right)  \right]
~\text{.}%
\end{align*}
The reader should notice our earlier result, that the related distributions
$\xi_{qn_{<N}}^{\left(  r\right)  }$ are all zero at zero temperature, merely
indicates that the respective $\eta_{qn_{<N}}^{\left(  r\right)  }$ are finite
(\textit{i.e.} neither zero nor divergent). The equations above contain no
driving terms, therefore the distributions should be constant functions of
$z$. In such a case we can make the replacement $G\rightarrow\delta\left(
z\right)  /2$ and performing the convolutions obtain a purely algebraic set of
equations,
\[
\left(  \eta_{qn_{<N}}^{\left(  r\right)  }\right)  ^{2}=\frac{\left(
1+\eta_{qn+1}^{\left(  r\right)  }\right)  \left(  1+\eta_{qn-1}^{\left(
r\right)  }\right)  }{\left(  1+\bar{\eta}_{qn}^{\left(  r+1\right)  }\right)
\left(  1+\bar{\eta}_{qn}^{\left(  r-1\right)  }\right)  }%
\]
that, is easy to verify, admits the solution
\[
\eta_{qn}^{\left(  r\right)  }=\frac{\sin\left(  \frac{n+r}{N+M}\pi\right)
\sin\left(  \frac{n+M-r}{N+M}\pi\right)  }{\sin\left(  \frac{r}{N+M}%
\pi\right)  \sin\left(  \frac{M-r}{N+M}\pi\right)  }-1~\text{.}%
\]
Plugging this result in the expression for the free energy we read off the
residual impurity entropy:
\[
S_{\mathrm{imp}}^{0}=\ln\frac{\sin\frac{\pi M}{N+M}}{\sin\frac{\pi}{N+M}}%
=\ln\frac{\sin\frac{\pi N}{N+M}}{\sin\frac{\pi}{N+M}}~\text{.}%
\]
This formula agrees with the one that was derived in the integral valence
limit\cite{jaz98} using the multi-channel version of the Coqblin-Schrieffer
model.\cite{cs69,nb80} That model is the Schrieffer-Wolff limit\cite{sw66} of
the multi-channel Anderson model that we are studying. We now turn to the
discussion of such limit.

\subsubsection{Integral Valence or Schrieffer-Wolff Limit\label{SW}}

In the standard approach --~at the level of the Hamiltonian~-- the
Schrieffer-Wolff (SW) limit is defined as a truncated unitary transformation
that eliminates the direct hybridization term and traces over (discards) the
less favorable states of valence from the impurity Hilbert space. This limit
leads to a projection onto a Hilbert subspace with the impurity site
permanently occupied by a \textit{local moment}. Starting from
$H_{\mathrm{MchA}}$, one obtains this way the multi-channel Coqblin-Schrieffer
(CS) model (the $N$-channel $SU\!\left(  M\right)  $ model or the $M$-channel
$SU\!\left(  N\right)  $ one depending on the sign of $\varepsilon$, see
below). In the following we will discuss how this effective limiting procedure
can be carried out at the level of the TBA equations.

Let us point out that the SW-transformation can be viewed as a first step in a
more detailed Renormalization Group analysis (which we will not carry out in
here). The multi-channel CS-model is therefore, when away from mixed valence,
the \textit{na\"{\i}ve} low temperature effective Hamiltonian of the full
multi-channel Anderson model. This effective Hamiltonian is relevant for the
description of the low energy dynamics for energy scales $E\ll T_{\mathrm{H}}%
$. In other words,
\[
H_{\mathrm{MchA}}\rightarrow H_{\mathrm{CS},x}+\left[  \mathcal{O\!\!}\left(
\frac{E}{T_{\mathrm{H}}}\right)  \text{ corrections in all sectors}\right]
\]
(with $x=s,q$ for the magnetic or quadrupolar sector, respectively). In the
limit of $\left\vert \varepsilon\right\vert \rightarrow+\infty$,
$T_{\mathrm{H}}\rightarrow+\infty$ and the projection is exact (albeit with a
vanishing exchange constant). For any finite $T_{\mathrm{H}}$, the effects of
the corrections, though `subleading' to those kept in $H_{\mathrm{CS},x}$, may
be important at low enough temperatures if they combine with other operators
to provide more singular contributions. This will be the case, for instance,
when the degeneracy of the lower-energy impurity configuration exceeds that of
the (degenerate) higher-energy one, as we shall stress in Sec.~\ref{END}. We
conclude therefore that the SW-transformation needs to be used with extreme
care in circumstances like the one described above.

We now turn to studying the integral valence limit via the Bethe-Ansatz. The
discussion of the integral valence limit turns out to be simpler using an
`inverted' set of TBA equations (trading back the recursions on the rank
indices for finite sums). In this alternative formulation, the equations read
(see the appendix for the kernel definitions),%
\begin{align*}
-\mathrm{\bar{f}}_{sn}^{\left(  r\right)  }=  &  R_{N}^{rs}\ast\left[
\mathrm{f}_{sn+1}^{\left(  s\right)  }+\hat{\delta}_{n,1}\mathrm{f}%
_{sn-1}^{\left(  s\right)  }+\delta_{n,1}\mathrm{\bar{f}}_{cr}-G^{-1}%
\ast\mathrm{f}_{sn}^{\left(  s\right)  }\right] \\
-\mathrm{\bar{f}}_{qn}^{\left(  r\right)  }=  &  -\delta_{n,N}~\tau^{-1}\Delta
E_{q}^{r}+\\
&  \left.
\begin{array}
[c]{l}%
+R_{M}^{rs}\ast\left[  \mathrm{f}_{qn+1}^{\left(  s\right)  }+\hat{\delta
}_{n,1}\mathrm{f}_{qn-1}^{\left(  s\right)  }-\right. \\
\qquad\qquad\qquad\qquad\left.  -G^{-1}\ast\mathrm{f}_{qn}^{\left(  s\right)
}-\delta_{n<N}\delta_{s,1}\mathrm{\bar{f}}_{cn}\right]
\end{array}
\right. \\
-\mathrm{f}_{cn_{<N}}=  &  -\tau^{-1}\Delta E_{s}^{n}+\\
&  \qquad\qquad+R_{N}^{nm}\ast\left[  \mathrm{\bar{f}}_{qm}^{\left(  1\right)
}+\mathrm{f}_{s1}^{\left(  m\right)  }-G^{-1}\ast\mathrm{\bar{f}}_{cm_{<N}%
}\right]
\end{align*}
where we defined the effective \textit{driving terms} (we remind the reader
that $\tau=\pi T/4\Delta$),%
\begin{align*}
\Delta E_{q}^{r}/\tau &  \equiv R_{M}^{r1}\ast\mathrm{\bar{f}}_{cN}\\
\Delta E_{s}^{n}/\tau &  \equiv R_{N}^{nN-1}\ast\mathrm{f}_{cN}~\text{.}%
\end{align*}

We need to discuss separately the quadrupolar and the magnetic scenarios. The
calculations are very similar in both cases. We turn our attention to the
former. In the quadrupolar SW limit ($\varepsilon_{q}\ll\varepsilon_{s}%
$),\footnote{We remind the reader that, for the sake of simplicity, much of
the discussion assumes $\mu=0$.} the $SU\!\left(  N\right)  \otimes
SU\!\left(  M\right)  $ Anderson model maps into an $N$-channel CS-model for a
local $SU\!\left(  M\right)  $ quadrupolar impurity. The key to understand
this limit is the study of the driving terms.

At low temperatures --~in a sense that will become precise below~-- and for
$\left\vert \lambda\right\vert \gg1$ (\textit{i.e.} away from the intrinsic
mixed-valence region) we can make the following approximation --~motivated by
our zero temperature solution for the ground state~-- that captures the
leading functional dependence in temperature and rapidity of the distribution
associated to the maximal charge-spin bound states: $\mathrm{f}_{cN}%
\simeq\operatorname{lex}\left(  M\lambda/\tau\right)  ~\theta\left(
-\lambda\right)  +\operatorname{lex}\left(  N\lambda/\tau\right)
~\theta\left(  \lambda\right)  $. The effective driving terms can then be
approximated as (for $\lambda\gg1$ in the quadrupolar case),%
\begin{align*}
\Delta E_{q}^{r}  &  \simeq-R_{M}^{r1}\ast\left[  M\lambda\right]
^{-}\underset{\lambda\gg1}{\longrightarrow}\frac{M^{2}}{4\pi}\sin\frac
{\pi\left(  M-r\right)  }{M}e^{-\frac{2}{M}\lambda}\\
\Delta E_{s}^{n}  &  \simeq R_{N}^{nN-1}\ast\left[  N\lambda\right]
^{+}\underset{\lambda\gg1}{\longrightarrow}n\lambda~\text{.}%
\end{align*}
As the temperature is lowered, both driving terms diverge. But since we have
that $\Delta E_{s}^{n}\gg\Delta E_{q}^{r}$ due to the different dependence in
rapidity, $\Delta E_{s}^{n}/\tau$ will drive to zero the distributions for the
charge rapidities ($\mathrm{f}_{cn_{<N}}$) faster than $\Delta E_{q}^{r}/\tau$
will drive the $\mathrm{f}_{qN}^{\left(  r\right)  }$'s. That way the
$\mathrm{f}_{qn}^{\left(  r\right)  }$'s are effectively cut away from the
other distributions (in this case the $\mathrm{f}_{sn}^{\left(  r\right)  }%
$'s) and they alone determine the impurity thermodynamics. The free energy is
given by
\[
F_{\mathrm{imp}}/\tau\simeq\varepsilon_{q}+\mu_{q}-\frac{1}{2\pi}\sum_{t}%
\int\frac{\sin\frac{\pi t}{M}}{\cos\frac{\pi t}{M}+\cosh\xi}\mathrm{f}%
_{q1}^{\left(  t\right)  }\left(  \xi\right)  ~d\xi
\]
where we performed the change of variables $M\xi=2\pi/J-2\lambda$ and
redefined the $\mathrm{f}_{q1}^{\left(  t\right)  }$'s as functions of the new
variable $\xi$. The goal of this change of variables is to remove the
\textit{coupling constant} ($J$) dependence from the expression of the free
energy and move it into the TBA equations. The only explicit dependence on $J$
will be in the driving terms and the same is true for the explicit temperature
dependence; this fact will allow us to identify the natural temperature scales
of the system. Remark that, in terms of the new variables, the main
contribution to the free energy comes from the value of the distributions
around $\xi\approx0$.\bigskip

\paragraph{High Temperature Scale:}

Consider the driving terms $\Delta E_{s}^{n}$. Changing variables we obtain,
\[
\Delta E_{s}^{n}/\tau\underset{\lambda\gg1}{\longrightarrow}n\lambda
/\tau=\left(  \frac{n\pi}{J}-\frac{nM}{2}\xi\right)  /\tau\approx\left(
\frac{n\pi}{J}\right)  /\tau~\text{.}%
\]
We shall thus define the Schottky temperature scale (the name will find its
motivation later, with the discussion of the specific heat):%
\[
\tau_{\mathrm{S}}\equiv\frac{N\pi}{2J}\Longrightarrow T_{\mathrm{S}}%
=\frac{N\Delta}{J}~\text{.}%
\]
For $\tau<\tau_{\mathrm{S}}$ (and $J<1$), the valence fluctuations are
quenched and the model goes into a regime where an effective description in
terms of a CS-model becomes appropriate. The distributions associated to the
charge rapidities go to zero and isolate the ones associated to the
quadrupolar rapidities that form a system of TBA equations identical to the
one of the CS-model. Notice that, as the temperature is lowered, these driving
terms diverge faster than the other ones that have to overcome a decaying
exponential in the numerator (cf.~with the caveats about the SW-transformation
discussed above).\bigskip

\paragraph{Low Temperature Scale:}

Consider now the driving terms $\Delta E_{q}^{r}$. Changing variables once
more we obtain
\[
\Delta E_{q}^{r}/\tau\underset{\lambda\gg1}{\longrightarrow}\frac{2}{N}%
\sin\frac{\pi\left(  M-r\right)  }{M}e^{\xi-\ln\left(  \tau/\tau_{\mathrm{k}%
}\right)  }%
\]
where%
\[
\tau_{\mathrm{K}}\equiv\frac{NM^{2}}{8\pi}e^{-\frac{2\pi}{MJ}}\Longrightarrow
T_{\mathrm{K}}=N\Delta\left(  \frac{M}{2\pi}\right)  ^{2}e^{-\frac{2\pi}{MJ}%
}~\text{.}%
\]
We have chosen to leave a factor of $2/N$ outside of the definition of the
Kondo scale in order to have a complete resemblance between the resulting TBA
equations and those for the scaling limit of a multi-channel
CS-model\cite{jaz98} (that CS-model, together with a cut-off prescription
$D_{\mathrm{eff}}=NM^{2}\Delta/4\pi^{2}$, is therefore the appropriate low
energy effective theory). Since the equations match, all the analysis done for
that model (finding the leading exponents of the specific heat coefficient and
susceptibilities, etc.) applies in this limit and we do not need to repeat
those considerations here.\cite{jaz98}\bigskip

\paragraph{The two Scales:\label{2scales}}

For the \textit{magnetic moment limit} ($\varepsilon_{s}\ll\varepsilon_{q}$)
we would find again two scales with the roles of $M$ and $N$ interchanged (we
omit the details since the considerations are very similar). As a shorthand,
we can extrapolate the two low energy scales (quadrupolar and magnetic) into
the high energy scales of the other regime (for the opposite sign of
$\varepsilon$). This serves the extra purpose of providing an \textit{ad hoc}
interpolation between both regimes and across the intrinsic mixed valence
region. We can define thus the two scales,\footnote{Notice that $T_{s}\neq
T_{\mathrm{S}}$; these two scales only coincide, asymptotically, in the
quadrupolar moment limit ($\varepsilon_{s}\gg\varepsilon_{q}$).}%
\begin{align*}
T_{q} &  \equiv\frac{2\Delta}{\pi}\tau_{q}\approx\frac{2\Delta}{\pi}\left(
M/2\right)  ^{2}\ln\left(  1+\frac{N}{2\pi}e^{-\frac{2\pi}{MJ}}\right)  \\
&  ~\\
T_{s} &  \equiv\frac{2\Delta}{\pi}\tau_{s}\approx\frac{2\Delta}{\pi}\left(
N/2\right)  ^{2}\ln\left(  1+\frac{M}{2\pi}e^{\frac{2\pi}{NJ}}\right)
~\text{.}%
\end{align*}

The role of these two temperature scales was illustrated in the schematic
picture of Fig.~\ref{teps}. In terms of them we can further write,
\begin{align*}
T_{\mathrm{H}}\left(  \varepsilon\right)   &  \equiv\max\left\{  T_{s}\left(
\varepsilon\right)  ,T_{q}\left(  \varepsilon\right)  \right\}  \underset
{\left\vert \varepsilon\right\vert \gg\Delta}{\longrightarrow}T_{\mathrm{S}%
}\left(  \varepsilon\right) \\
&  ~\\
T_{\mathrm{L}}\left(  \varepsilon\right)   &  \equiv\min\left\{  T_{s}\left(
\varepsilon\right)  ,T_{q}\left(  \varepsilon\right)  \right\}  \underset
{\left\vert \varepsilon\right\vert \gg\Delta}{\longrightarrow}T_{\mathrm{K}%
}\left(  \varepsilon\right)
\end{align*}
that serve as boundaries among the high-, intermediate-, and low-temperature
regimes. At high temperatures the impurity is in a mixed valence state. As the
temperature is lowered, the system crosses the first temperature scale
indicated as $T_{\mathrm{H}}$, for $\left\vert \varepsilon\right\vert
\gtrsim\Delta$ this coincides with $T_{\mathrm{S}}$ (the chemical potential is
taken to be zero to lighten the notation). At this point the system enters a
regime that can be approximately described with the respective CS-model. At
first, the system is in the `unscreened' local moment regime, extending
between $T_{\mathrm{H}}$ and the lower scale $T_{\mathrm{L}}=T_{\mathrm{K}}$.
The larger the value of $\left\vert \varepsilon\right\vert $ the wider the
temperature window of this regime. For $\left\vert \varepsilon\right\vert
\lesssim\Delta$ the two energy scales merge, indicating that there is no
moment formation. As the temperature is lowered beyond $T_{\mathrm{L}}$, the
moment screening commences and the physics is asymptotically governed by a
line of fixed points parametrized by the value of $\varepsilon/\Delta$ (or
some observable that depends on it and varies along the line, as for instance
the charge valence of the impurity). The different points in the line share
the same value for the impurity entropy, and the same set of leading exponents
of the specific heat coefficient (\textit{i.e.}~$\gamma=C_{\mathrm{imp}}/T$)
or the different susceptibilities. However, the prefactors of the different
leading terms will in general vary along the line and could in principle be
determined by direct measurement. Thus a multi-channel Kondo effect takes
place for any $\varepsilon$ as the temperature is lowered. It is amusing to
note that for $\varepsilon=0$, in particular, the Kondo effect takes place
\textit{without} moment formation.

\subsubsection{Asymptotic values}

In this section we will be interested in finding the values that the different
distributions (in this case the $\eta$'s) take when $\lambda\rightarrow
\pm\infty$. This corresponds to the limit of infinite $\left\vert
\varepsilon-\mu\right\vert $ at \textit{finite temperature}. These results
will be required in the next section, where the numerical solution of the TBA
equations is discussed. Remark that, due to the behavior of the two energy
scales derived above, in the limits considered in this section one always has
$T_{\mathrm{k}}\lll T\lll T_{\mathrm{s}}$ and the $\lambda\rightarrow\pm
\infty$ limits shall be, respectively, identified with the \textit{infinite}
temperature limits of the effective quadrupolar and magnetic multi-channel CS-models.

From the results above and inspection of the equations, the reader can
convince himself that,
\[
\left\{
\begin{array}
[c]{c}%
\left\{
\begin{array}
[c]{c}%
\lim\limits_{\lambda\rightarrow+\infty}\mathrm{f}_{cN}=+\infty\\
~\\
\lim\limits_{\lambda\rightarrow+\infty}\mathrm{\bar{f}}_{cN}=0
\end{array}
\right. \\
~\\
\left\{
\begin{array}
[c]{c}%
\lim\limits_{\lambda\rightarrow-\infty}\mathrm{f}_{cN}=0\\
~\\
\lim\limits_{\lambda\rightarrow-\infty}\mathrm{\bar{f}}_{cN}=+\infty~\text{.}%
\end{array}
\right.
\end{array}
\right.
\]
We shall use this result to analyze the TBA equations in the asymptotic limit.
The two limits, $\lambda\rightarrow\pm\infty$, are different and we shall
consider them separately.\bigskip

\paragraph{Right Asymptotics:}

Let us first study the limit of $\lambda\rightarrow+\infty$ that we call the
right asymptotics of the distributions. We want to find the values
$\overset{\rightharpoonup}{\eta}_{xn}^{\left(  r\right)  }=\lim
\limits_{\lambda\rightarrow+\infty}\eta_{xn}^{\left(  r\right)  }$ that the
different distributions take in the limit (with $x=s,q$). Except for
$\eta_{cN}$ that is unbounded, all the other $\eta$'s acquire finite values
and for $\lambda$ very large can be taken as constants. From the limit
$\mathrm{f}_{cN}\underset{\lambda\rightarrow+\infty}{\longrightarrow}+\infty$
it follows that all the $\overset{\rightharpoonup}{\bar{\eta}}_{cn_{<N}}$'s go
to zero and the equations for the $\overset{\rightharpoonup}{\eta}%
_{sn}^{\left(  r\right)  }$'s and those for the $\overset{\rightharpoonup
}{\eta}_{qn}^{\left(  r\right)  }$'s form two identical sets decoupled from
each other and obeying the following recurrence relation:%
\[
\overset{\rightharpoonup}{\eta}_{xn+1}^{\left(  r\right)  }+1=\left(
1+\overset{\rightharpoonup}{\bar{\eta}}_{xn}^{\left(  r+1\right)  }\right)
\frac{\left(  \overset{\rightharpoonup}{\eta}_{xn}^{\left(  r\right)
}\right)  ^{2}}{\left(  1+\overset{\rightharpoonup}{\eta}_{xn-1}^{\left(
r\right)  }\right)  }\left(  1+\overset{\rightharpoonup}{\bar{\eta}}%
_{xn}^{\left(  r-1\right)  }\right)  ~\text{.}%
\]
In the case of uniform or `Zeeman' splitting, all the $\alpha_{x}^{r}$'s
entering the large $n$ asymptotic condition are equal and an analytic solution
for the asymptotic values is known. It is easy to show with some algebra that
the solutions are:\cite{ad84,dr87,tsvelik87,jaz98}
\begin{align*}
\overset{\rightharpoonup}{\eta}_{xn}^{\left(  r\right)  }+1  &  =\frac
{\sinh\left(  \left(  n+r\right)  \alpha_{x}\right)  \sinh\left(  \left(
n+N_{x}-r\right)  \alpha_{x}\right)  }{\sinh\left(  r\alpha_{x}\right)
\sinh\left(  \left(  N_{x}-r\right)  \alpha_{x}\right)  }\\
&  \underset{\alpha_{x}\rightarrow0}{\longrightarrow}\frac{\left(  n+r\right)
\left(  n+N_{x}-r\right)  }{r\left(  N_{x}-r\right)  }%
\end{align*}
where $N_{s}=N$, $N_{q}=M$ and $\alpha_{x}$ lost its dependence on the rank
index. A closed analytic solution is not known for the case of a more general
\textit{crystal field splitting} (more general splittings might be relevant to
make the connection with the experimental systems that motivated the model,
this point will be discussed further in the next section). Notice that when
the values of the $\overset{\rightharpoonup}{\eta}$'s are required for a
particular splitting, they can always be found numerically.\bigskip

\paragraph{Left Asymptotics:}

We study now the opposite limit of $\lambda\rightarrow-\infty$, in order to
find the asymptotic values $\overset{\leftharpoonup}{\eta}_{xn}^{\left(
r\right)  }=\lim\limits_{\lambda\rightarrow-\infty}\eta_{xn}^{\left(
r\right)  }$. Except for $\bar{\eta}_{cN}$ that is unbounded, all the other
$\eta$'s acquire finite values and for $\lambda$ very large can be taken as
constants. From the limit $\mathrm{\bar{f}}_{cN}\underset{\lambda
\rightarrow-\infty}{\longrightarrow}+\infty$ it follows that all the
$\overset{\leftharpoonup}{\eta}_{qN}^{\left(  r\right)  }$'s go to zero and
the equations for the remaining distributions form two identical sets
decoupled from each other and similar to those of the right asymptotic limit.
The solution of that case can be applied to this one if the following
identifications are made (cf. with the discussion of Fig.~\ref{mchtbas}):
\[
\left\{
\begin{array}
[c]{ccc}%
\overset{\leftharpoonup}{\eta}_{sn}^{\left(  r\right)  } & = & \overset
{\rightharpoonup}{\eta}_{sM+n}^{\left(  r\right)  }\\
& ~ & \\
\overset{\leftharpoonup}{\bar{\eta}}_{cn_{<N}} & = & \overset{\rightharpoonup
}{\eta}_{sM}^{\left(  n\right)  }\\
& ~ & \\
\overset{\leftharpoonup}{\bar{\eta}}_{qn_{<N}}^{\left(  r\right)  } & = &
\overset{\rightharpoonup}{\eta}_{sM-r}^{\left(  n\right)  }\\
& ~ & \\
\overset{\leftharpoonup}{\eta}_{qN}^{\left(  r\right)  } & = & 0\\
& ~ & \\
\overset{\leftharpoonup}{\eta}_{qn_{>N}}^{\left(  r\right)  } & = &
\overset{\rightharpoonup}{\eta}_{qn-N}^{\left(  r\right)  }~\text{.}%
\end{array}
\right.
\]
These are the same identifications that would be required for an explicit
discussion of the Schrieffer-Wolff limit in the local magnetic moment regime,
($\varepsilon_{s}\ll\varepsilon_{q}$), in order to match the resulting TBA
equations to those of the corresponding magnetic multi-channel CS-model.

Once the right and left asymptotic limits are both known, one is then ready to
tackle the problem of solving the TBA equations numerically; as we proceed to
explain in the latter parts of this section.

\subsection{Numerical Solution}

In order to access the thermodynamics of the system at all temperatures, we
shall resort to the numerical solution of the TBA equations. This task was
carried out in the past for other integrable impurity
models.\cite{rajan83,dr85,jaz98,cz99,zcja02} The equations of the
multi-channel Anderson impurity model have many similarities with those
considered previously for other types of impurities, but they present also
important differences. We give below a brief outline of the numerical
procedure that we developed followed by the results we thus obtained.

\subsubsection{Outline of the Numerical Procedure}

We solve the TBA equations using a standard iterative scheme first introduced
in the work of Rajan.\cite{rajan83} The idea is to use the TBA equations in
their recursive formulation, start with some educated initial guess, and
iterate them until certain convergence criterion is met (this general scheme
is sometimes known as Kepler's method; cf.~Ref.~%
[\onlinecite{ts02}]%
). Our particular implementation borrows ideas mainly from the previous work
by Costi and Zar\'{a}nd for the anisotropic Kondo model.\cite{cz99} In that
work, the closure of the infinite set of TBA equations into a finite set
brings in a great simplification. This is not, however, the case in general
and we need to address the issue of truncation of the infinite hierarchy of
recursions. The standard procedure is to define some boundary levels (in our
case $\mathrm{f}_{sn_{s}+1}^{\left(  r\right)  }$ and $\mathrm{f}_{qn_{q}%
+1}^{\left(  r\right)  }$, for suitably large values $n_{s}$ and $n_{q}$ of
the level index) and fix them. It was done in the past by taking those
boundary distributions as constants equal to the average of their right and
left asymptotic values that are known analytically.\cite{rajan83,ts02} This
approximation is good when $n_{s}$ and $n_{q}$ are fairly large (depending on
the particular model). In the case at hand the TBA equations are more
complicated than those for the Kondo model (in a way that will become clear
below) and we cannot afford the computational cost of taking too large values
for $n_{s}$ and $n_{q}$. The alternative we found was to start with some
educated guess for the distributions up to $\mathrm{f}_{xn_{x}+1}^{\left(
r\right)  }$ (with $x=s,q$) that interpolates smoothly between the right and
left asymptotic values (discussed above) respecting the monotonicity
properties of the solution. Then we iterate the equations to find new values
for the different distributions up to the levels $n=n_{s},n_{q}$. At the end
of each iteration we update the value of the boundary distributions
($n=n_{s}+1,n_{q}+1$) using the $\mathrm{f}_{xn_{x}}^{\left(  r\right)  }$'s
suitably shifted and rescaled to match the analytically known asymptotic
values corresponding to the boundary levels. This procedure is repeated until
stable convergence is achieved on the free energy function that we recalculate
at each iteration.

The rapidity dependence of the distributions in the case of the Kondo TBA
equations is associated to variations in temperature, and the two asymptotic
limits correspond to the zero and infinite temperature limits.\cite{afl83}
This is in contradistinction to the case of the TBA equations for the
multi-channel Anderson model, for which the rapidity dependence is related to
variations of the coupling ($\sim1/J$). We will therefore require modest
computational effort to compute the free energy for different values of the
coupling and determine quantities like charge susceptibilities, but
temperature dependence will require independent runs for each value of
temperature required. Since determining temperature dependence is essential,
our computational task becomes typically two to three orders of magnitude
larger than for Kondo impurities; depending on the range and number of
temperature points desired and not counting the inherent extra complexity of
our TBA equations.

Once we have a finite number of equations involving a finite number of
continuous distributions, we need to discretize those distributions. This is
conceptually done in two steps, the first one being the introduction of
cut-offs on the rapidity axis. We have to choose large right and left cut-offs
that enter once the distributions are approximately constant functions of the
rapidity reaching their respective right and left asymptotic values. Second,
we need to discretize the axis interval between the cut-offs. This is done
defining three regions. First a small region centered around zero that we
discretize using a fine mesh. The size of this region is chosen depending on
the parameters of the problem so that it encloses all the intervals where
different distributions show rapid variations. This typically happens for
region boundary values of the rapidities such that the magnitude of their
associated coupling corresponds to a Kondo temperature of the order of the
temperature set for the system. Second, the two regions to the right and left
of the central one are discretized logarithmically until reaching the cut-offs
defined above. In these shoulder regions all the distributions vary slowly as
they attain their asymptotic values.%

\begin{figure*}[htb]
\begin{center}%
{\includegraphics[
height=4.7253in,
width=6.9332in
]%
{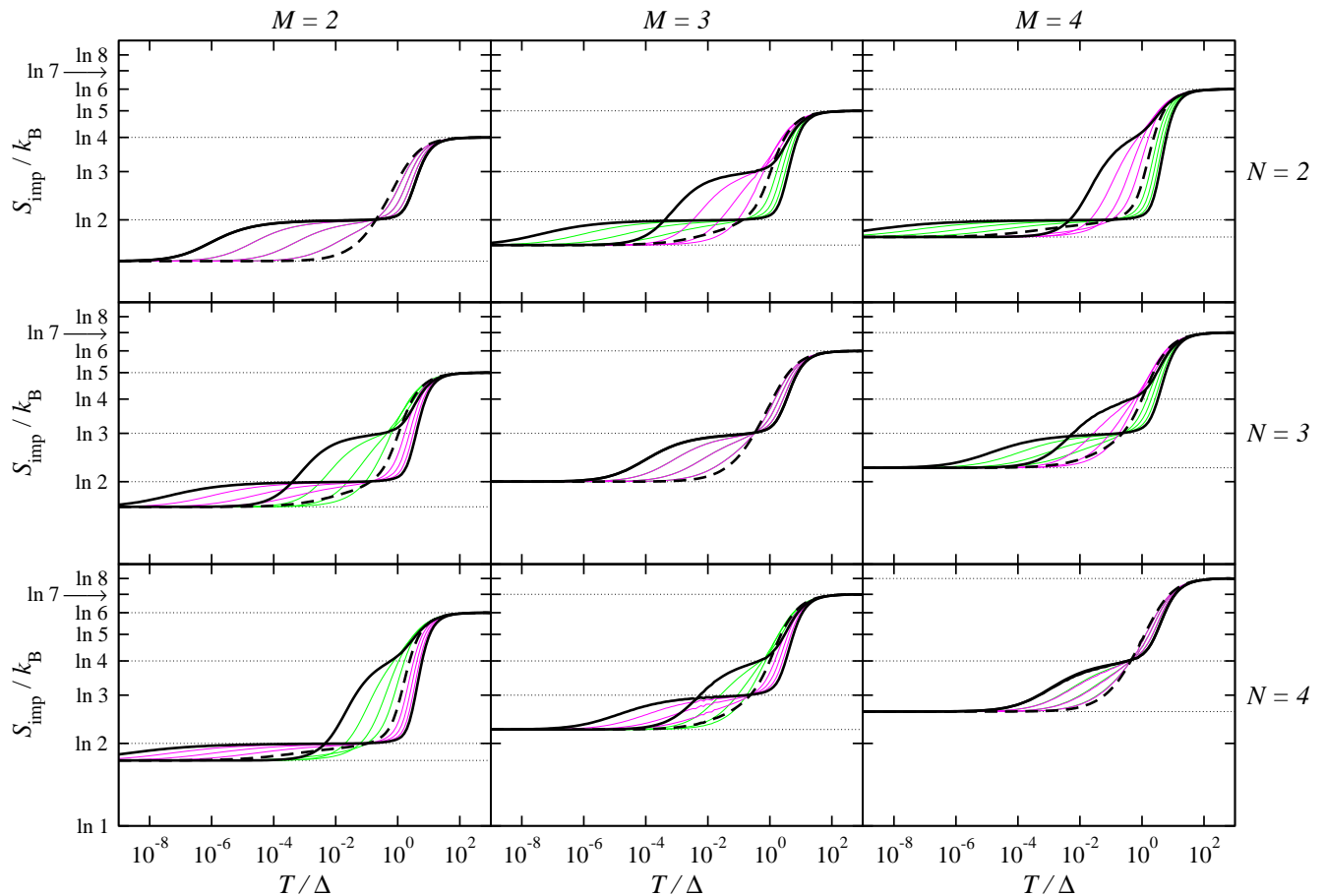}%
}%
\caption{\label{entropy9}Impurity entropy as a function of
temperature. The different curves correspond to:
$\varepsilon/\Delta=0$ (dashed lines); $\varepsilon/\Delta =\pm2$,
$\pm4$, $\pm6$ (light lines); and $\varepsilon/\Delta=\pm8$ (dark
solid lines). The different panels give the results for different
values of $N$ and $M$ as indicated to the right and above,
respectively. Curves for different signs of $\varepsilon$ are
degenerate in the diagonal panels. In the off-diagonal ones they
can be identified from the value of the entropy in the
intermediate plateaux (see text).}%
\end{center}
\end{figure*}%
%

\begin{figure*}[htb]
\begin{center}%
{\includegraphics[
height=4.7253in,
width=6.7957in
]%
{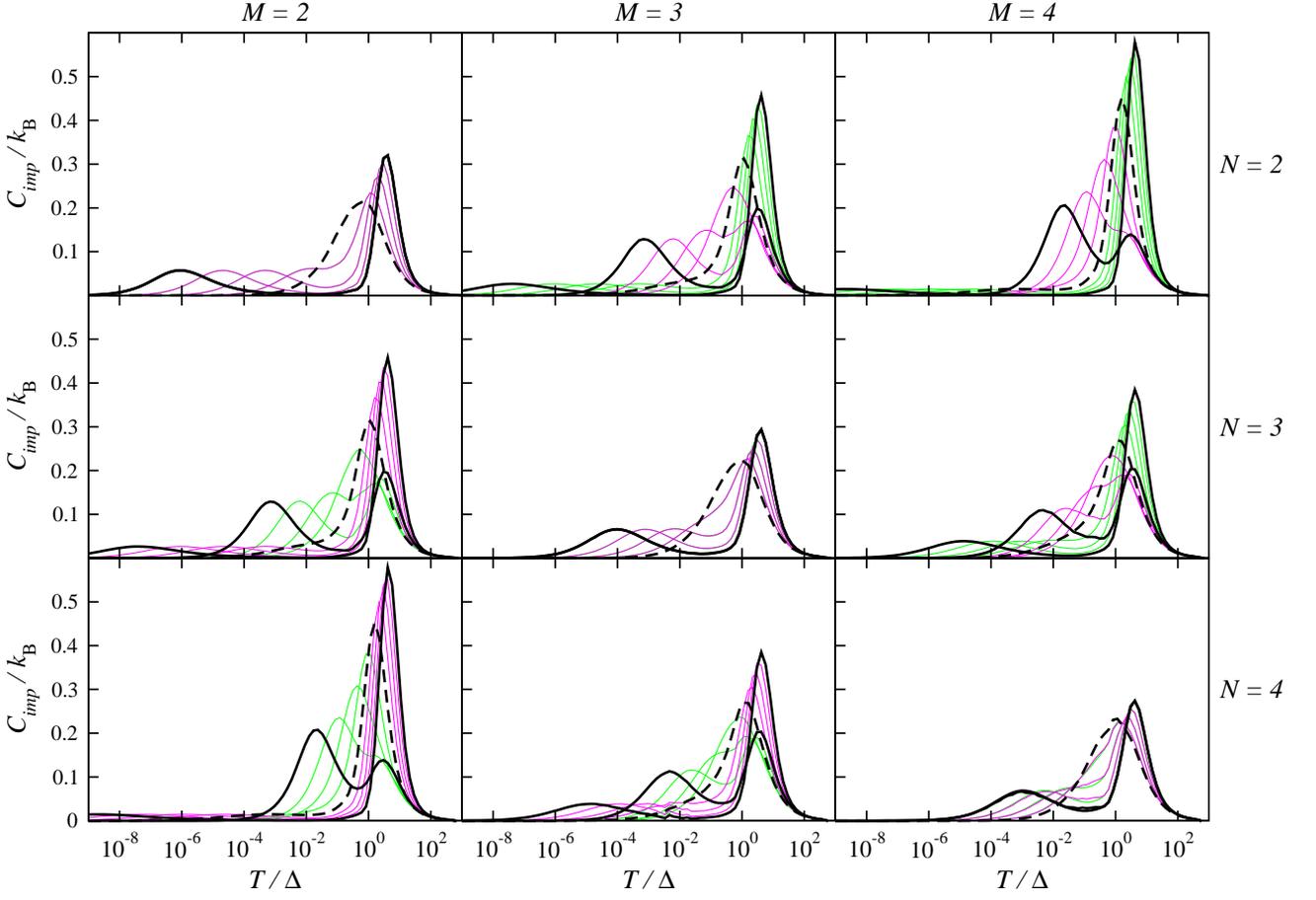}%
}%
\caption{\label{cv9}Impurity specific heat as a function of
temperature. The different curves correspond to:
$\varepsilon/\Delta=0$ (dashed lines); $\varepsilon /\Delta=\pm2$,
$\pm4$, $\pm6$ (light lines); and $\varepsilon/\Delta=\pm8$ (dark
solid lines). The different panels give the results for different
values of $N$ and $M$ as indicated to the right and above,
respectively. Curves for different signs of $\varepsilon$ can be
identified from the location of the
Kondo anomaly or by comparison with the entropy plots.}%
\end{center}
\end{figure*}%
%

\begin{figure*}[htb]
\begin{center}%
{\includegraphics[
height=4.7253in,
width=6.7568in
]%
{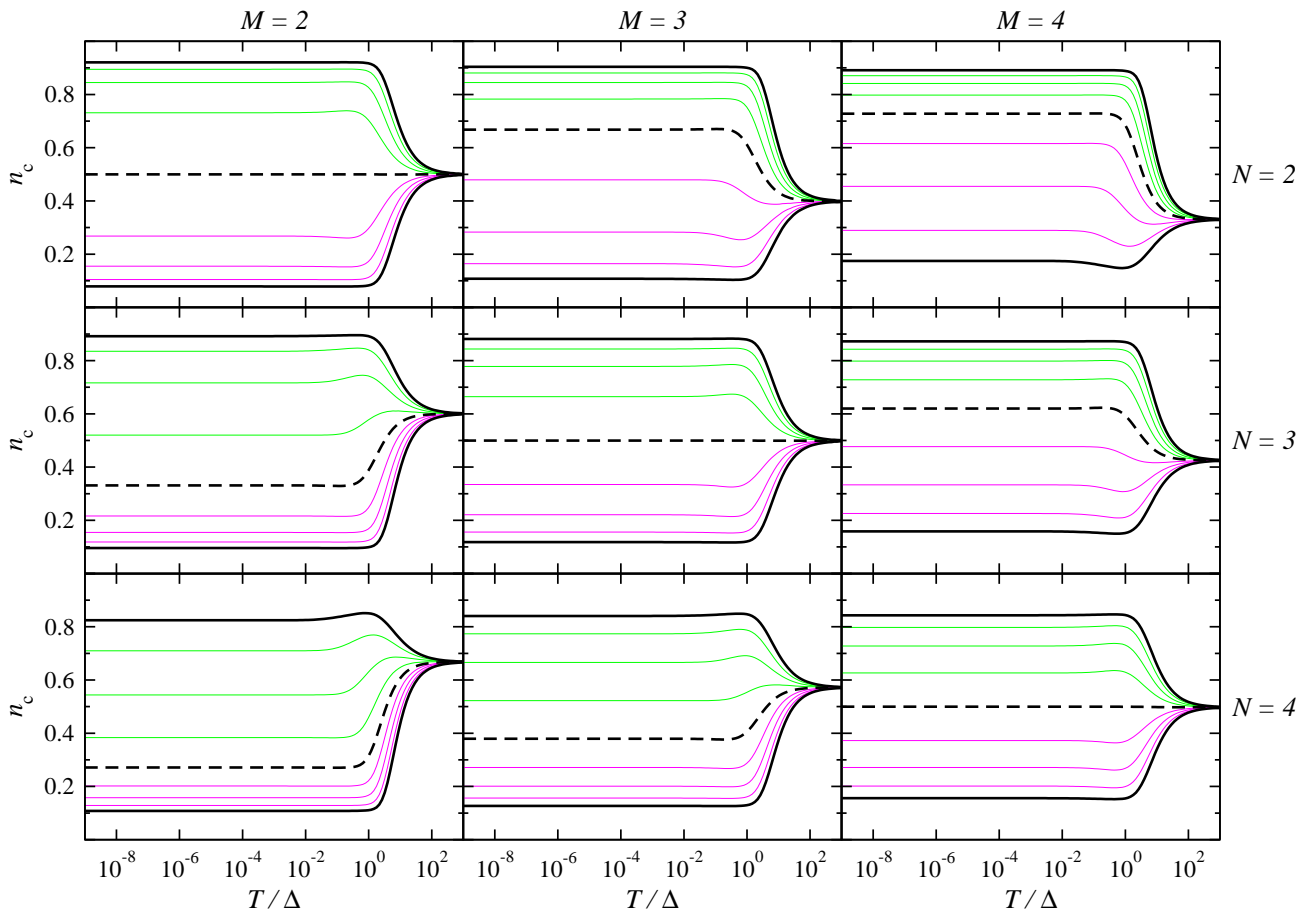}%
}%
\caption{\label{nc9}Impurity charge valence as a function of
temperature. The different curves correspond to:
$\varepsilon/\Delta=0$ (dashed lines); $\varepsilon /\Delta=\pm2$,
$\pm4$, $\pm6$ (light lines); and $\varepsilon/\Delta=\pm8$ (dark
solid lines). The different panels give the results for different
values of $N$ and $M$ as indicated to the right and above,
respectively. Curves for $\varepsilon$ positive or negative are,
respectively, below or above the
$\varepsilon=0$ line.}%
\end{center}
\end{figure*}%

The different convolution kernels required in the calculations should be
evaluated, this is done a single time at the beginning of the calculations
when they are stored as matrices. The kernels are relatively rapidly varying
functions of the rapidities as compared with the TBA distributions, so we
discretize them using a denser mesh. We use a mesh that is locally $n$ times
(typically $n\sim10$) finer than the one used for the TBA distributions and
these are interpolated to the points in the finer mesh using cubic splines.
The extra points in the finer mesh are chosen to be the $n$-th order
quadrature points of each subinterval of the coarser one. The convolution
integrals can thus be split into the different subintervals and carried out
with high precision in each one of them using Gaussian quadrature (corrections
for the introduction of the right and left cut-offs are also implemented).
Since accuracy is very important for certain applications,\cite{jab03} the
evaluation of the different kernels requires a high precision implementation
of the sums involving four digamma functions of complex arguments (see
appendix). To carry it out we wrote an algorithm based on `a precision
approximation of the gamma function' due to Lanczos.\cite{lanczos64}

With everything set to start the iterative evaluation of the TBA equations,
one last point is worth mentioning. The evaluation of the right-hand side of
these equations requires the implementation of the ubiquitous function
$\operatorname{lex}\left(  x\right)  $. This implementation is, however,
unsurprisingly subtle. We first notice that, since $\operatorname{lex}\left(
x\right)  =\left[  x\right]  ^{+}+\operatorname{lex}\left(  -\left\vert
x\right\vert \right)  $, it suffices to implement the function for negative
arguments. This we do by first evaluating $e^{-\left\vert x\right\vert }$ and
then using a careful implementation of $\ln\left(  1+x\right)  $ with the
property of cancellation of rounding errors.\cite{Higham} This algorithm
ensures monotonicity, machine precision accuracy, and `graceful underflow'.

We show below some of the results that were obtained with different
implementations of the described numerical procedure.

\subsubsection{Numerical Results}

As stated above, one of the important issues that makes the numerical analysis
of the TBA equations required is the study of the temperature dependence in
its full range. We give below illustrative results of this dependence where
the presence of the two cross-over scales discussed already on analytic
grounds is clearly observed. We start by showing results for the model in the
absence of crystalline or external applied fields.

In Fig.~\ref{entropy9} we show several plots of the entropy as a function of
temperature. The different panels correspond to different symmetries of the
model; $N=2,3,4$ row-wise and $M=2,3,4$ column-wise as indicated. The
different curves in each panel correspond to different values of the energy
splitting between the two impurity configurations ($\varepsilon=\varepsilon
_{s}-\varepsilon_{q}$) and, without loss of generality, the total chemical
potential is taken to be zero.\footnote{Alternatively, the whole discussion
can be rephrased in terms of $\varepsilon-\mu$.} In all the panels, the dark
solid lines correspond to $\varepsilon/\Delta=\pm8$ and the dark dashed line
corresponds to the \textit{extreme} mixed valence case of $\varepsilon=0$ (or
$\varepsilon=\mu$). The remaining thin lines correspond to intermediate values
of the energy splitting ($\varepsilon/\Delta=\pm2,\pm4,\pm6$) and are given to
illustrate how the system \textit{interpolates} among the different limits of
large, positive and negative, and vanishing $\varepsilon$. On the one hand, in
the high-temperature limit, the entropy is in each case given by
$S_{\mathrm{imp}}^{\infty}\left(  N,M\right)  =k_{\mathrm{B}}\ln\left(
N+M\right)  $; as expected since $N+M$ is the total size of the impurity
Hilbert space. On the other hand, in the low-temperature limit, the value of
the entropy tends to the $S_{\mathrm{imp}}^{0}\left(  N,M\right)  $ values we
found analytically (see Sec.~\ref{ResEntropy}). We remark that both these
limiting values are independent of $\varepsilon$.

For intermediate temperatures, the figure shows how the impurity entropy is
quenched from its high- to its low-temperature values as the temperature
decreases. This happens as a two stage process for large values of
$\varepsilon$ (solid lines) and as a single stage process in the mixed valence
case (dashed lines), all in accordance with the theoretical discussion given
above (see Fig.~\ref{teps} and the discussion in Sec.~ \ref{2scales}).
Following the solid lines as the temperature lowers, the first quenching stage
corresponds to the cross-over scale $T_{\mathrm{S}}$ and the impurity entropy
attains the values $k_{\mathrm{B}}\ln\left(  N\right)  $ or $k_{\mathrm{B}}%
\ln\left(  M\right)  $ for $\varepsilon_{s}\ll\varepsilon_{q}$ or
$\varepsilon_{s}\gg\varepsilon_{q}$ respectively. (As expected, in the case of
$N=M$, the curves for different positive and negative values of $\varepsilon$
are degenerate; on the other hand, for $N\neq M$, the curves for the same
absolute values but opposite signs of $\varepsilon$ are in precise
correspondence upon exchange of the values of $N$ and $M$).\footnote{The
different panels were nevertheless computed all independently as a check of
the correctness of the program.} These intermediate-regime plateaux correspond
to the formation of a free local moment when $\left\vert \varepsilon
\right\vert \gg\Delta$. As the temperature lowers further, following always
the solid lines, the systems reach the Kondo cross-over scale, $T_{\mathrm{K}%
}$, below which the entropy tends to the universal values $S_{\mathrm{imp}%
}^{0}\left(  N,M\right)  $ characteristic of the different infrared non-Fermi
liquid fixed points. This second quenching stage takes place as electrons in
the different channels compete to screen the local moment of the intermediate
regime. It is this dynamical \textit{frustrated screening} process that is
responsible for the non-trivial nature of the fixed point. In the mixed
valence case, exemplified by the dashed line, the two cross-over energy scales
are of comparable magnitude and as a result the quenching of the entropy as
the temperature is lowered happens on a single stage. Remarkably, the same
limiting value of the impurity entropy is found also in this case; all
consistent with the picture given above, that the infrared physics is governed
by a line of fixed points (Sec.~\ref{2scales}). In a previous work, we
explored this explicitly (in the $N=M=2$ case) using Boundary Conformal Field
Theory and showed that the different points in the line are connected by an
exactly marginal operator.\cite{jab03}\bigskip

The results for the impurity contribution to the specific heat --~obtained
upon differentiation of the impurity entropy~-- are shown in Fig.~\ref{cv9}.
The different panels, and the different lines in each of them, follow the same
conventions as in the entropy plots. In accordance with the generic two steps
shape of the entropy (degenerate in the mixed valence case), the specific heat
shows a general two humps structure (again degenerate for mixed valence). The
lower temperature one is sometimes referred to in the experimental literature
as the `Kondo anomaly', and its location is correspondingly given by
$T_{\mathrm{K}}$. On the other hand, the higher temperature one is referred to
as the `Schottky anomaly' and its position is given by $T_{\mathrm{S}}$. There
is of course no new information in this figure as compared with the previous
one, but it is the specific heat rather than the entropy or the free energy
the quantity that is most often `directly' accessible in the
experiments.\cite{rietschel88}\bigskip

It is illustrative to look as well at the behavior as a function of
temperature of the impurity charge valence, $n_{c,\mathrm{imp}}=\sum_{\sigma
}\left\langle f_{\sigma}^{\dagger}f_{\sigma}\right\rangle $. This is provided
in Fig.~\ref{nc9} following identical conventions as in the plots for the
entropy and the specific heat. As expected, in the high-temperature limit, the
impurity valence approaches the values,%
\[
n_{c,\mathrm{imp}}^{\infty}\left(  N,M\right)  =\frac{N}{N+M}%
\]
corresponding to the impurity Hilbert space fraction with `magnetic
character'. This is an expression of the fact that, in that limit, all the
impurity states are equiprobable. Subsequently, as the temperature crosses the
value $T_{\mathrm{S}}$, the impurity charge valence changes and approaches
rapidly what will be its zero temperature value, $n_{c,\mathrm{imp}}%
^{0}\left(  N,M\right)  $. This change is at the origin of the Schottky
anomaly in the specific heat. As it should be, the impurity approaches integer
valence when $\left\vert \varepsilon\right\vert \gg\Delta$. The charge valence
goes to zero in the \textit{quadrupolar} limit of large and positive
$\varepsilon$, and goes to one in the opposite, \textit{magnetic} limit, of
$\varepsilon$ large but negative. The values of $n_{c,\mathrm{imp}}^{0}$ for
intermediate energy splittings are difficult to calculate and would constitute
a good non-trivial test for approximate theories like those based on $\left(
1/N\right)  $-expansions. Careful comparisons of NCA and NRG were carried out
this way in the single-channel case,\cite{ckw96} but for the multi-channel
case the NRG calculations rapidly become very demanding for present day
computational resources and this kind of comparisons were not done. Also,
since the full cross-over takes place at the highest energy scale, the
convergence of $n_{c,\mathrm{imp}}^{0}$ should be relatively fast in NRG
calculations. Thus in the future it might constitute a useful observable to
monitor the progress of such computations by comparing with the exact solution.

As a check, we verified numerically that%
\[
\left.  n_{c,\mathrm{imp}}^{0}\left(  N,M\right)  \right\vert _{\varepsilon
}+\left.  n_{c,\mathrm{imp}}^{0}\left(  M,N\right)  \right\vert _{-\varepsilon
}=1
\]
so that, in particular, one finds that $\left.  n_{c,\mathrm{imp}}^{0}\left(
N,N\right)  \right\vert _{\varepsilon=0}=1/2$. The analytic form of
$n_{c,\mathrm{imp}}^{0}\left(  N,M\right)  $ can be computed using the zero
temperature results of Sec.~\ref{Tzero} (with the caveats given there for the
case of $N\neq M$). For instance, in the two-channel case, we have%
\begin{align*}
n_{c,\mathrm{imp}}^{0}\left(  2,2\right)  =  &  ~\frac{2}{\pi}\int_{-\infty
}^{+\infty}\int_{-\infty}^{0}\frac{z~dz}{\cosh\left[  \frac{\pi}{2\Delta
}\left(  x-z\right)  \right]  }\times\\
&  ~\qquad\qquad\times\frac{\mu-\varepsilon+x}{\left[  \left(  \mu
-\varepsilon+x\right)  ^{2}+4\Delta^{2}\right]  ^{2}}~dx\\
&  ~\\
\longrightarrow &  \left\{
\begin{array}
[c]{ccc}%
\approx1 & \qquad\text{for}\qquad & \varepsilon\ll-\Delta\\
& ~ & \\
=\frac{1}{2} & \qquad\text{for}\qquad & \varepsilon=0\\
& ~ & \\
\approx0 & \qquad\text{for}\qquad & \varepsilon\gg\Delta
\end{array}
\right.
\end{align*}
(where we reintroduced the chemical potential). The inner integral is related
to $\xi_{q2}$ and has an exact expression in terms of dilogarithms. It is
interesting to observe how, for certain values of the energy splittings, the
charge valence first increases and then decreases as a function of the
lowering temperature (or \textit{vice versa}) as it goes from
$n_{c,\mathrm{imp}}^{\infty}$ to $n_{c,\mathrm{imp}}^{0}$. This behavior is
already present in the case of single-channel degenerate Anderson
impurities.\cite{ckw96}\bigskip

Before proceeding, let us add some remarks on the technical side. We
calculated both the entropy and the specific heat by numerically
differentiating the free energy computed using the TBA distributions obtained
with the algorithm described in the section above. Numerical differentiation
is a delicate procedure very susceptible to truncation and roundoff errors and
in general sensitive to any noise in the original data. The specific heat is,
since a second derivative is involved, rather sensitive to this type of errors
(the discretization of the temperature plays an interrelated role as well). In
the event of better accuracy being required, alternative ways of computing
derived thermodynamic quantities are possible. One common procedure is to set
up secondary sets of integral equations for the different derivatives of the
TBA distributions (see, for instance, Ref.~%
[\onlinecite{cz99}]%
). Once the original distributions are found, these equations can be solved to
determine their derivatives. Using them one can calculate directly (or with a
smaller number of numerical differentiations) the sought derived quantities.
The procedure to solve these auxiliary sets of equations will be in general
similar to that used for the original TBA equations; making the total
computational effort increase accordingly.\bigskip

\begin{figure*}[htb]
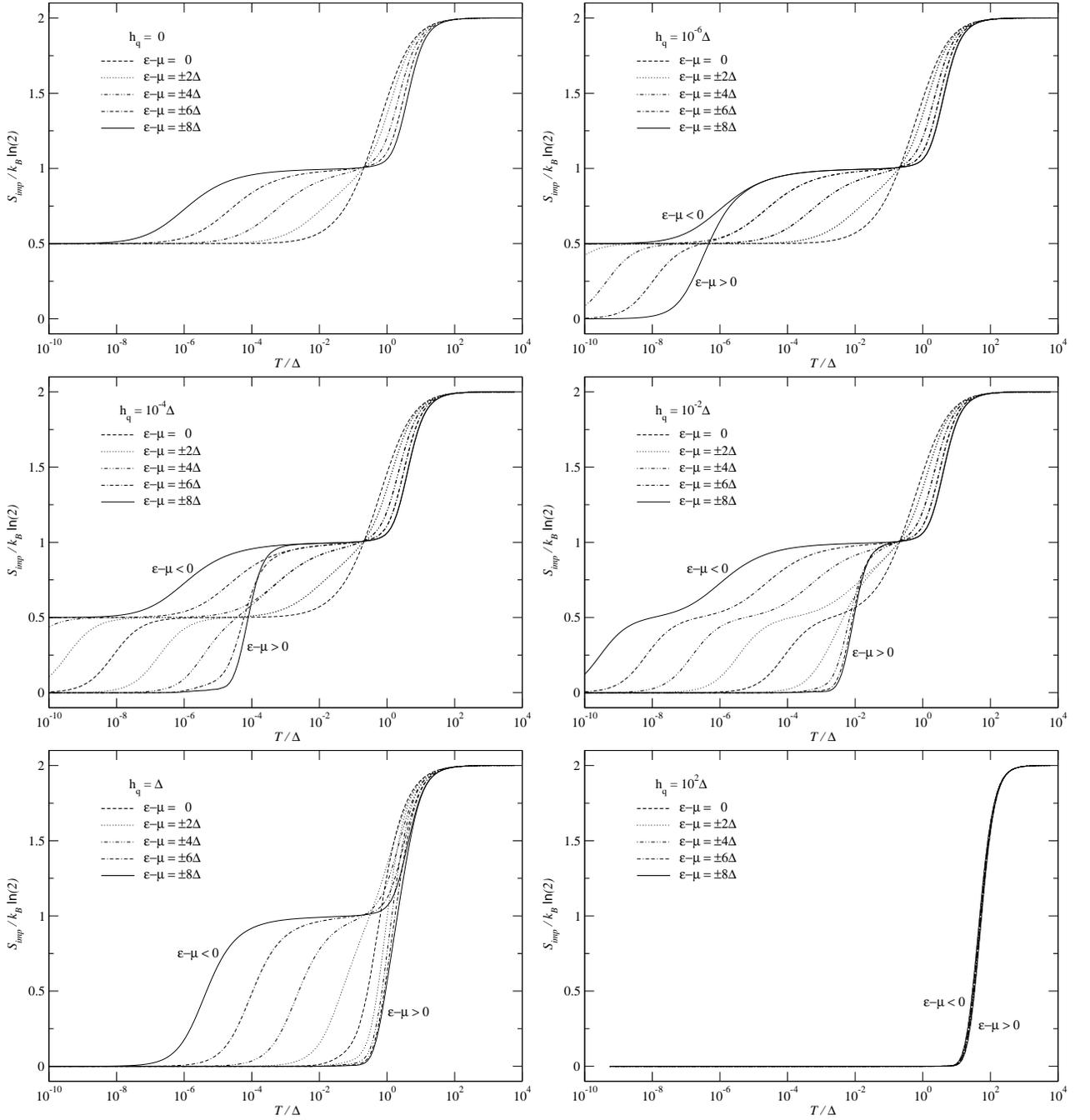

\begin{center}%
\begin{tabular}
[c]{ll}%
{\includegraphics[
height=2.3056in,
width=3.3096in
]%
{SvsT_Qzero.eps}%
}%
&
{\includegraphics[
height=2.3056in,
width=3.3096in
]%
{SvsT_Qem6.eps}%
}%
\\%
{\includegraphics[
height=2.3056in,
width=3.3096in
]%
{SvsT_Qem4.eps}%
}%
&
{\includegraphics[
height=2.3056in,
width=3.3096in
]%
{SvsT_Qem2.eps}%
}%
\\%
{\includegraphics[
height=2.3056in,
width=3.3096in
]%
{SvsT_Qem0.eps}%
}%
&
{\includegraphics[
height=2.3056in,
width=3.3096in
]%
{SvsT_Qep2.eps}%
}%
\end{tabular}
\caption[Entropy as a function of temperature for various applied
fields.]{\label{svst} Entropy as a function of temperature for
various applied fields (we have chosen to consider quadrupolar
flavor fields). The top-left panel corresponds to the zero field
case. To the left of it and following downwards the field
increases hundred-fold each time as
$h_q=10^{-6}\Delta,...,10^{2}\Delta$. The values of $\varepsilon$
are the same as in Fig.~\ref{entropy9} (the chemical potential was
reintroduced in the notation for this figure), and the top-left
panels of both figures are in correspondence.}
\end{center}
\end{figure*}

We turn now to discuss the effects of external fields on the physics of the
impurity. For the sake of simplicity we restrict ourselves to the two-channel
case ($N=M=2$). Since magnetic and quadrupolar fields are \textit{relevant}
perturbations, the presence of any of them has important effects on the
entropy. In fact these perturbations drive the system to a totally different
--~this time Fermi liquid~-- line of fixed points characterized by a zero
value of the residual impurity entropy (cf.~with the situation in the
single-channel case). This is illustrated in Fig.~\ref{svst} where the entropy
as a function of temperature is shown for different values of energy splitting
between doublets ($\varepsilon$) and quadrupolar field ($h_{q}$).%

In the upper left panel of the figure we reproduce again the results for zero
field. The curves with positive and negative $\varepsilon$ coincide in this
case (remember we are setting $\mu=0$ in this discussion). As we go from high
to low temperatures, the system crosses over from a state with entropy
$S_{\mathrm{imp}}=k_{\mathrm{B}}\ln4$ to one with entropy $S_{\mathrm{imp}%
}=k_{\mathrm{B}}\ln\sqrt{2}$. This second value is one of the hallmarks of the
non-trivial non-Fermi liquid fixed point that governs the low energy physics
of the model.\cite{ad84} In the next panel to the right we show the effect of
turning on an external field. For $h_{q}/\Delta=10^{-6}$ one observes that the
positive-$\varepsilon$ curves stay in an $S_{\mathrm{imp}}=k_{\mathrm{B}}%
\ln\sqrt{2}$ plateau for a very short temperature interval (that disappears
altogether when the impurity energy splitting is sufficiently large) and a
third quenching step of the entropy takes it to a zero value final state,
indicative of a Fermi liquid fixed point. As the reader can observe across the
different panels, the quenching to zero entropy of the large positive
$\varepsilon$ curves takes place when the temperature is lowered until it
becomes of the same order of magnitude as the applied field ($T\sim h_{q}$).

On the other hand, the fate of the negative-$\varepsilon$ curves becomes more
evident as we increase the field further (see the middle two panels). In this
case too, the non-Fermi liquid fixed point is unstable and the entropy goes to
zero, but this takes place at much lower temperatures than for positive
$\varepsilon$. What happens is that at low temperatures the splitting of the
higher multiplet (that plays the role of the orbital channel in the usual
multi-channel Kondo scenario) renders the model an effective spin-half
single-channel exchange model in which the impurity is `exactly' screened at
low temperatures and the fixed point is a Fermi liquid.\cite{sl95} Curves with
intermediate values of $\varepsilon$ interpolate continuously between these
two behaviors very much in the way they did in the other cases that we
discussed above.

As the field is increased and approaches the scale of the first stage of
entropy quenching (\textit{i.e.}~$T_{\mathrm{S}}$) and beyond, the
different-$\varepsilon$ curves start to collapse and the temperature of the
cross-over transition moves up following the field (compare the last two
panels in Fig.~\ref{svst}). The entropy is quenched in a single stage from
$S_{\mathrm{imp}}=k_{\mathrm{B}}\ln4$ to zero at the same time as the impurity
charge valence goes from $n_{c,\mathrm{imp}}=1/2$ to zero.

\section{Summary and Outlook\label{END}}

We have carried out a detailed analysis of the multi-channel Anderson impurity
model. We have demonstrated its integrability, discussed the details of the
Bethe-Ansatz solution, and shown a number of illustrative results for the
thermodynamics of the model.

\begin{figure}
[t]
\begin{center}
\includegraphics[
height=0.4238in,
width=3.1704in
]%
{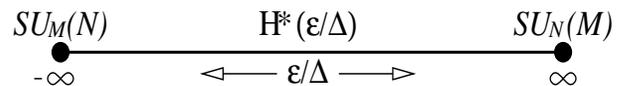}%
\caption{Schematic representation of the line of fixed points.}%
\label{line}%
\end{center}
\end{figure}

The Bethe-Ansatz allows the study of the model on all energy scales, we
proceed now to summarize our findings and to put them into context. We begin
by discussing the low-temperature regime and the characterization of the low
energy fixed point theory that corresponds to the microscopic model. We have
identified the existence of a line of boundary critical fixed points that
governs the low energy physics of the impurity (illustrated in Fig.~\ref{line}%
) and shown how it is related to the fixed point theory of the multi-channel
Coqblin-Schrieffer model. In fact, every point in the line describes non-Fermi
liquid physics. At the extreme ends ($\left\vert \varepsilon\right\vert
\rightarrow+\infty$), the fixed point theory corresponds to an $SU_{M}\left(
N\right)  $ or $SU_{N}\left(  M\right)  $ Hamiltonian (in the sense of BCFT).
For any finite value of $\varepsilon$, and in particular for all the values
that correspond to the mixed valence region, the physics is given by a line of
$SU_{M}\left(  N\right)  \oplus SU_{N}\left(  M\right)  $ theories. Along this
line of fixed points the behavior, for instance, of the specific heat would be
given by (cf. Ref.~%
[\onlinecite{jaz98}]%
),%
\[
C_{\mathrm{imp}}\underset{N\neq M}{\sim}\lambda_{c}^{2}\left(  \varepsilon
\right)  ~T+\lambda_{s}^{2}\left(  \varepsilon\right)  ~T^{\frac{2N}{N+M}%
}+\lambda_{q}^{2}\left(  \varepsilon\right)  ~T^{\frac{2M}{N+M}}%
\]
($N=M$ corresponding to the marginal case when both spin and quadrupolar
sectors contribute $T\ln T$ leading temperature dependences; cf.~Ref.~%
[\onlinecite{jab03}]%
). The smallest exponent dominates as $T\rightarrow0$, except possibly at the
limit when $\varepsilon\rightarrow\pm\infty$ and $\lambda_{s}^{2}\left(
\varepsilon\right)  $ or $\lambda_{q}^{2}\left(  \varepsilon\right)  $ vanish,
respectively. Thus these two limits do not commute when $f<n$ for the
$SU_{f}\left(  n\right)  $ end of the fixed points line (with $n,f=N,M>1$).
Notice therefore, that as $\varepsilon$ is varied along the line at small but
finite temperature, the observed critical behavior will vary accordingly.

The same finite-$T$ cross-over in non-Fermi liquid character (magnetic
\textit{vs}.~quadrupolar) would manifest itself in a comparative study of both
susceptibilities, whose expected leading low-temperature behaviors, after
subtraction of asymptotic temperature-independent contributions, are:
\[
\left\{
\begin{array}
[c]{c}%
\chi_{\mathrm{imp}}^{s}-\chi_{\mathrm{0}}^{s}\underset{N\neq M}{\sim}%
\lambda_{s}^{2}\left(  \varepsilon\right)  ~T^{\frac{N-M}{N+M}}\\
\chi_{\mathrm{imp}}^{q}-\chi_{\mathrm{0}}^{q}\underset{N\neq M}{\sim}%
\lambda_{q}^{2}\left(  \varepsilon\right)  ~T^{\frac{M-N}{N+M}}%
\end{array}
\right.
\]
(here $N=M$ is again the marginal case, for which $\ln T$ dependencies are
expected). Set for instance $\varepsilon$ in the quadrupolar regime, then the
corresponding $\lambda_{q}^{2}\left(  \varepsilon\right)  $ is large while
$\lambda_{s}^{2}\left(  \varepsilon\right)  $ can be arbitrarily small.
However, one will still find that, for all finite values of $\varepsilon$, the
spin susceptibility will eventually dominate over the quadrupolar one if it
happens to carry the singular exponent (\textit{i.e.} for $N<M$).

Note that the level with the lower degeneracy always dictates the physics at
sufficiently low temperatures even if it is a very high-energy level. This may
be surprising from the point of view of the Schrieffer-Wolff limit (see
Sec.~\ref{SW}). Naively, one would expect that the effects of the
energetically unfavorable level could be simply integrated out. This is not
always the case. When the degeneracy of this level is the smaller one, it will
end up dominating at sufficiently low $T$ (as long as $|\varepsilon|$ remains
finite).\footnote{Technically, in our case the SW transformation needs to be
carried out carefully to the next order to see that the operators associated
with the high-energy level mix with the low-energy ones and yield singular
contributions when their associated degeneracy is the lower one.} In other
words, discarding the energetically unfavorable yet less degenerate
configuration also eliminates the frustration that is induced in the bulk
electrons when they try to screen the virtual moment of such an excited state;
but it is this frustration that would have been responsible for the appearance
of singular exponents in the impurity thermodynamics.

This picture might help shed light upon some of the unexplained and sometimes
contradictory results observed in the experiments.\cite{cox98,stewart01} In
particular, interesting possibilities are open up for better understanding of
the intermediate-to-low temperature phase of certain heavy fermion compounds,
inespecially those believed to be near mixed valence (the list is rather
large, for an example see below).%

Thus far we discussed the low-temperature regime (FP --~dominated by the line
of \textit{ fixed points}). We have also studied in detail the other regimes:
the high-temperature valence fluctuation regime (FV) and the intermediate
local moment regimes (LMM and LQM) ---~the initialisms refer to the names used
in Fig.~\ref{teps}. We identified the two energy scales ($T_{s}$ and $T_{q}$)
associated with the spin and quadrupolar degrees of freedom, that cross each
other in the intermediate valence region ($\left\vert \varepsilon\right\vert
\lesssim\Delta$) and interchange roles as the high-temperature ($T_{\mathrm{H}%
}$) and low-temperature ($T_{\mathrm{L}}$) scales that indicate the transition
zones among different regimes. In the intermediate valence region the two
scales `coincide', indicating the direct transition between the high- and the
low-temperature behaviors. The system never develops a local moment, but goes
directly into the low-temperature, multi-channel Kondo like, non-Fermi liquid
phase governed by the corresponding fixed points. In all the cases,
\textit{i.e.}~for all values of $\varepsilon$, the fixed points are unstable
to the application of an external field acting either on the magnetic or the
quadrupolar degrees of freedom.

In subsequent work we intend to use the results presented here to analyze the
different multi-channel scenarios for certain heavy-fermion compounds like
\textrm{U}$_{1-x}$\textrm{Th}$_{x}$\textrm{Be}$_{13}$.\cite{kc99,hbtgbcs01} It
was shown that the two-channel Anderson model is not sufficient to account for
the relevant number of impurity degrees of freedom required to explain the
available specific heat measurements.\cite{ba02} More complicated impurity
models, like for instance an $SU\!\left(  2\right)  \otimes SU\!\left(
5\right)  $ model in the presence of crystal field splittings, hold a
considerable promise in that respect.\cite{ba02,ab03}

\begin{acknowledgments}
During the different stages of this work, we enjoyed discussions with
F.~B.~Anders, P.~Coleman, T.~Costi, T.~Giamarchi, A.~Jerez, H.~Johannesson,
H.~R.~Krishnamurthy, J.~Kroha, A.~Rosch, A.~Schiller, N.~Shah, and
P.~W\"{o}lfle. One of the authors was partly supported by the Swiss National
Science Foundation through MaNEP.
\end{acknowledgments}

\appendix

\section{Quantum Inverse Scattering Method\label{appendix_QISM}}

In the following we will skip the details (and assume them known) of the
standard case of models with $SU\!\left(  2\right)  $ internal symmetry, and
refer the reader to the presentation given in the ICTP lectures by one of the
authors.\cite{Andrei} We will adopt a notation very similar, though not
identical, to the one in those lectures.\footnote{The differences stemming
from the changed definition of the monodromy matrix: it will be the first
index of the scattering matrices, rather than the second one, that shall be
used as auxiliary space.} We will discuss directly the case of $SU\!\left(
3\right)  $ symmetry conveniently generalized to the situation when both
`particles' and `antiparticles' are present. This generalization is required
to solve the eigenvalue problem of the multi-channel Anderson model (the
standard \textit{Nested Bethe-Ansatz} (NBA) as discussed in the literature,
see for instance Ref.~%
[\onlinecite{tw83}]%
, does not suit this case), and was done before only for $SU\!\left(
3\right)  $ spin chains.\cite{ar96,ar97} Finally, we will close the appendix
with the generalization to the case of $SU\!\left(  N\right)  $ symmetry.

\subsection{Models with $SU\!\left(  3\right)  $ Symmetry, Impurities, and
Periodic Boundary Conditions}

Let us consider an integrable model with $SU\!\left(  3\right)  $ symmetric
scattering matrices given by,
\begin{align*}
S_{jn\neq0}  &  =\frac{\left(  \alpha_{j}-\alpha_{n}\right)  \mathbf{I}%
_{jn}+ic\mathbf{P}_{jn}}{\left(  \alpha_{j}-\alpha_{n}\right)  +ic}%
=S_{jn}\left(  \alpha_{j}-\alpha_{n}\right) \\
&  ~\\
S_{j0}  &  =\mathbf{I}_{j0}+\frac{e^{-i\delta\left(  \alpha_{j}-\alpha
_{0}\right)  }-1}{3}\mathbf{Q}_{j0}=S_{j0}\left(  \alpha_{j}-\alpha
_{0}\right)  ~\text{.}%
\end{align*}
We choose to consider a case where $S_{j0}$ has a different structure than
$S_{jn\neq0}$. In particular we consider the kind of impurity S-matrix that
arises in the flavor sector of the three-channel Anderson model. The
parameters {$\alpha_{j},\alpha_{0}$} are arbitrary at this point, and will be
chosen later to be those that specify the multi-channel Anderson model. The
eigenvalue problem given in terms of the above S-matrices is not tractable
with the standard NBA formalism; we develop below the required extensions.

\subsubsection{Monodromy Matrix}

As a first step we define a matrix that captures the monodromy conditions of
our eigenvalue problem and use its transfer matrix to rewrite the problem:
\[
\Xi_{A}\left(  \alpha\right)  =S_{AN}\left(  \alpha-\alpha_{N}\right)  \ldots
S_{A1}\left(  \alpha-\alpha_{1}\right)  S_{A0}\left(  \alpha-\alpha
_{0}\right)
\]
(this is equivalent to adding an auxiliary extra particle $A=N+1$). The
purpose of the auxiliary space will be to allow us to conveniently organize
the products of the S-matrices (see below). We define the \textit{transfer
matrix} as:
\[
T\left(  \alpha\right)  =\operatorname*{tr}\nolimits_{A}\Xi_{A}\left(
\alpha\right)
\]
where $\operatorname*{tr}\nolimits_{A}$ denotes taking the trace in the
auxiliary space $A$, and it follows that
\[
T\left(  \alpha_{j}\right)  =Z_{j}~\text{.}%
\]

As the amplitudes $\vec{A}$ should be simultaneous eigenvectors of all the
eigenvalue problems $Z_{j}$, it is necessary and sufficient that $\left[
Z_{j},Z_{l}\right]  =0$. We can go further and require $\left[  T\left(
\alpha\right)  ,T\left(  \beta\right)  \right]  =0$ for all values of $\alpha$
and $\beta$. This is guaranteed if there exists a matrix $R$ such that,
\[
R_{AB}\left(  \alpha-\beta\right)  \Xi_{A}\left(  \alpha\right)  \Xi
_{B}\left(  \beta\right)  =\Xi_{B}\left(  \beta\right)  \Xi_{A}\left(
\alpha\right)  R_{AB}\left(  \alpha-\beta\right)
\]
we will refer to this identity as the \textit{fundamental commutation
relation}\emph{ }(FCR).

\subsubsection{Fundamental Commutation Relations}

By repeated application of the Yang-Baxter relations, it can be shown that the
matrix
\[
R_{AB}\left(  \alpha\right)  =S_{AB}\left(  \alpha\right)  =\frac
{\alpha\mathbf{I}_{AB}+ic\mathbf{P}_{AB}}{\alpha+ic}%
\]
satisfies the FCR. If we write down the expression for the monodromy matrix
explicitly in the auxiliary space, we have (using a notation that is
convenient for the Nested Bethe-Ansatz):
\begin{align*}
\Xi\left(  \alpha\right)   &  =\left(
\begin{array}
[c]{ccc}%
A\left(  \alpha\right)  & B^{2}\left(  \alpha\right)  & B^{3}\left(
\alpha\right) \\
C^{2}\left(  \alpha\right)  & D^{22}\left(  \alpha\right)  & D^{23}\left(
\alpha\right) \\
C^{3}\left(  \alpha\right)  & D^{32}\left(  \alpha\right)  & D^{33}\left(
\alpha\right)
\end{array}
\right) \\
&  ~\\
T\left(  \alpha\right)   &  =A\left(  \alpha\right)  +D^{22}\left(
\alpha\right)  +D^{33}\left(  \alpha\right)  ~\text{.}%
\end{align*}
Fully expanding the FCR in the two auxiliary spaces, it can be seen, after
some algebra, that the sub-matrix $D$ verifies the corresponding FCR of the
$SU\!\left(  2\right)  $ case,
\begin{multline*}
R_{AB}^{\left(  2\right)  }\left(  \alpha-\beta\right)  D_{A}\left(
\alpha\right)  D_{B}\left(  \beta\right)  =\\
=D_{B}\left(  \beta\right)  D_{A}\left(  \alpha\right)  R_{AB}^{\left(
2\right)  }\left(  \alpha-\beta\right)
\end{multline*}
with
\[
R_{AB}^{\left(  2\right)  }\left(  \alpha\right)  =S_{AB}^{\left(  2\right)
}\left(  \alpha\right)  =\frac{\alpha\mathbf{I}_{AB}^{\left(  2\right)
}+ic\mathbf{P}_{AB}^{\left(  2\right)  }}{\alpha+ic}~\text{.}%
\]
This means that Yang-Baxter is obeyed by the sub-matrix. Defining $u_{\alpha
}=\frac{\alpha-ic}{\alpha}$ and $v_{\alpha}=\frac{ic}{\alpha}$, one can write
the following commutation relations between different components of the
monodromy matrix:\footnote{For reasons of brevity we shall henceforth use
either notation: $A_{\alpha}\equiv A\left(  \alpha\right)  $, $u_{\alpha
}\equiv u\left(  \alpha\right)  $, etc.}
\begin{align*}
A_{\alpha}B_{\beta}^{s}  &  =u_{\alpha-\beta}B_{\beta}^{s}A_{\alpha}%
+v_{\alpha-\beta}B_{\alpha}^{s}A_{\beta}\\
D_{\alpha}B_{\beta}^{s}  &  =u_{\beta-\alpha}\;B_{\beta}^{s}D_{\alpha}%
S_{As}^{\left(  2\right)  }+v_{\beta-\alpha}\mathbf{\;}B_{\alpha}^{s}D_{\beta
}\mathbf{P}_{As}^{\left(  2\right)  }\\
B_{\beta}^{s}B_{\beta^{\prime}}^{s^{\prime}}  &  =B_{\beta^{\prime}%
}^{s^{\prime}}B_{\beta}^{s}S_{ss^{\prime}}^{\left(  2\right)  }~\text{.}%
\end{align*}
They will be used extensively in what follows.

\subsubsection{Reference State or Pseudo-vacuum}

We use the following local basis for the individual Hilbert space of each
particle (where the `anti-particle' one is for the impurity Hilbert Space):
\[
\left\vert u\right\rangle ,\left\vert \bar{u}\right\rangle =\left(
\begin{array}
[c]{c}%
1\\
0\\
0
\end{array}
\right)  ;\quad\left\vert d\right\rangle ,\left\vert \bar{d}\right\rangle
=\left(
\begin{array}
[c]{c}%
0\\
1\\
0
\end{array}
\right)  ;\quad\left\vert s\right\rangle ,\left\vert \bar{s}\right\rangle
=\left(
\begin{array}
[c]{c}%
0\\
0\\
1
\end{array}
\right)
\]
and we define the following \textit{highest weight reference state }which we
will use to construct the Fock Space for the problem:
\[
\left\vert \omega\right\rangle =\left(  \bigotimes_{n=1}^{N}\left\vert
u\right\rangle _{n}\right)  \otimes\left\vert \bar{s}\right\rangle
_{0}~\text{.}%
\]

Using the definitions $a_{\alpha}=\frac{\alpha}{\alpha+ic}$ and $b_{\alpha
}=\frac{ic}{\alpha+ic}$ we can write down explicitly the scattering matrices
in auxiliary space and see how they act on the reference state. Using that
information we find for the monodromy matrix:
\[
\Xi\left(  \alpha\right)  \left\vert \omega\right\rangle =\left(
\begin{array}
[c]{ccc}%
\left\vert \omega\right\rangle  & \ast & \ast\\
0 & \Delta^{\prime\prime}\left(  \alpha\right)  \left\vert \omega\right\rangle
& \ast\\
0 & 0 & \Delta^{\prime\prime\prime}\left(  \alpha\right)  \left\vert
\omega\right\rangle
\end{array}
\right)
\]
with
\begin{align*}
\Delta^{\prime\prime}\left(  \alpha\right)   &  =\prod_{j=1}^{N}a\left(
\alpha-\alpha_{j}\right) \\
\Delta^{\prime\prime\prime}\left(  \alpha\right)   &  =\frac{2+e^{-i\delta
\left(  \alpha-\alpha_{0}\right)  }}{3}\Delta^{\prime\prime}\left(
\alpha\right)  ~\text{.}%
\end{align*}
Notice, for later use, that $\Delta^{\prime\prime}\left(  \alpha_{j}\right)
=0$. We define $\Delta\left(  \alpha\right)  =\Delta^{\prime\prime}\left(
\alpha\right)  +\Delta^{\prime\prime\prime}\left(  \alpha\right)  $. For the
transfer matrix we find (verifying that the reference state is indeed an
eigenstate of it),
\[
T_{\alpha}\left\vert \omega\right\rangle =\left(  A_{\alpha}+D_{\alpha}%
^{22}+D_{\alpha}^{33}\right)  \left\vert \omega\right\rangle =\left(
1+\Delta_{\alpha}\right)  \left\vert \omega\right\rangle ~\text{.}%
\]

\subsubsection{Descendant States}

We will construct descendant eigenstates from the reference eigenstate that
constitutes the highest weight state of the largest possible representation.
Let us consider a state obtained by acting with the linear combination of $M$
\textit{flavor-lowering} operators (where $X_{s_{1}\ldots s_{M}}$ is an
arbitrary tensor):
\[
\left\vert \vec{\beta}\right\rangle =\sum_{\left\{  s_{i}\right\}  }B^{s_{1}%
}\left(  \beta_{1}\right)  \ldots B^{s_{M}}\left(  \beta_{M}\right)
\;X_{s_{1}\ldots s_{M}}\left\vert \omega\right\rangle ~\text{.}%
\]
Using the FCR we find (the nomenclature of \textit{wanted} and
\textit{unwanted} terms is standard, and the same as in Ref.~%
[\onlinecite{Andrei}]%
),%
\begin{align*}
\left.  A_{\alpha}\left\vert \vec{\beta}\right\rangle \right\vert
_{\text{wanted}}=  &  \left(  \prod_{n=1}^{M}u_{\alpha-\beta_{n}}\right)
\left\vert \vec{\beta}\right\rangle \\
&  ~\\
\left.  D_{\alpha}\left\vert \vec{\beta}\right\rangle \right\vert
_{\text{wanted}}=  &  \left(  \prod_{n=1}^{M}u_{\beta_{n}-\alpha}\right)
\times\\
\times B_{\beta_{1}}^{s_{1}}\ldots &  B_{\beta_{M}}^{s_{M}}\;D_{\alpha
}\;S_{As_{M}}^{\left(  2\right)  }\ldots S_{As_{1}}^{\left(  2\right)
}\;X_{s_{1}\ldots s_{M}}\left\vert \omega\right\rangle \\
&  ~\\
\left.  \operatorname*{tr}\nolimits_{A}D_{\alpha}\left\vert \vec{\beta
}\right\rangle \right\vert _{\text{wanted}}=  &  \left(  \prod_{n=1}%
^{M}u_{\beta_{n}-\alpha}\right)  \times\\
&  \times B_{\beta_{1}}^{s_{1}}\ldots B_{\beta_{M}}^{s_{M}}\;T_{\alpha
}^{\left(  2\right)  }\;X_{s_{1}\ldots s_{M}}\left\vert \omega\right\rangle
\end{align*}
where in the last line we made the following definitions:
\[
\Xi^{\left(  2\right)  }\left(  \alpha\right)  =D_{\alpha}\;S_{As_{M}%
}^{\left(  2\right)  }\ldots S_{As_{1}}^{\left(  2\right)  }%
\]%
\[
T^{\left(  2\right)  }\left(  \alpha\right)  =\operatorname*{tr}%
\nolimits_{A}\Xi^{\left(  2\right)  }\left(  \alpha\right)  ~\text{.}%
\]
Let us define the \textit{reduced} monodromy matrix,
\begin{multline*}
\tilde{\Xi}^{\left(  2\right)  }\left(  \alpha\right)  =S_{As_{M}}^{\left(
2\right)  }\left(  \alpha-\beta_{M}\right)  \ldots S_{As_{1}}^{\left(
2\right)  }\left(  \alpha-\beta_{1}\right)  \equiv\\
\equiv\left(
\begin{array}
[c]{cc}%
\tilde{A}\left(  \alpha\right)  & \tilde{B}\left(  \alpha\right) \\
\tilde{C}\left(  \alpha\right)  & \tilde{D}\left(  \alpha\right)
\end{array}
\right)
\end{multline*}
written in the \textit{reduced} auxiliary space, and whose elements act only
on the space of indexes $\left\{  s_{i}\right\}  $ (\textit{i.e.}~that spanned
by all the possible $X_{s_{1}\ldots s_{M}}$). Notice that the elements of
$\tilde{\Xi}^{\left(  2\right)  }$ and those of $D$ commute with each other.
We write down their `combined' product explicitly:
\[
\Xi_{\alpha}^{\left(  2\right)  }=D_{\alpha}\tilde{\Xi}_{\alpha}^{\left(
2\right)  }=\left(
\begin{array}
[c]{cc}%
A_{\alpha}^{\left(  2\right)  } & B_{\alpha}^{\left(  2\right)  }\\
C_{\alpha}^{\left(  2\right)  } & D_{\alpha}^{\left(  2\right)  }%
\end{array}
\right)  ~\text{.}%
\]
Since both $D$ and $\tilde{\Xi}^{\left(  2\right)  }$ satisfy the Yang-Baxter
relations, we have a new set of FCR that are obeyed:
\begin{multline*}
R_{AB}^{\left(  2\right)  }\left(  \alpha-\beta\right)  \Xi_{A}^{\left(
2\right)  }\left(  \alpha\right)  \Xi_{B}^{\left(  2\right)  }\left(
\beta\right)  =\\
=\Xi_{B}^{\left(  2\right)  }\left(  \beta\right)  \Xi_{A}^{\left(  2\right)
}\left(  \alpha\right)  R_{AB}^{\left(  2\right)  }\left(  \alpha
-\beta\right)  ~\text{.}%
\end{multline*}

The eigenvalue we are seeking involves more than a particular realization of
the auxiliary tensor $X_{s_{1}\ldots s_{M}}$ (cf. with the NBA formalism). We
can define a highest weight reference state for the space of $\left\{
s_{i}\right\}  $ in the standard way (notice that $\tilde{\Xi}^{\left(
2\right)  }$ is the usual monodromy matrix that appears in the $SU\!\left(
2\right)  $ case). We therefore consider the `combined' reference state
\[
\left\vert \omega\right\rangle _{2}=\left\vert \tilde{\omega}\right\rangle
\left\vert \omega\right\rangle \qquad\text{with}\qquad\left\vert \tilde
{\omega}\right\rangle =\bigotimes_{n=1}^{M}\left\vert \uparrow\right\rangle
_{n}%
\]
given by the direct product of a reference state in the space of indexes and
$\left\vert \omega\right\rangle $ the previously defined reference state. In
particular we have:
\[
\tilde{\Xi}_{\alpha}^{\left(  2\right)  }\left\vert \tilde{\omega
}\right\rangle =\left(
\begin{array}
[c]{cc}%
\left\vert \tilde{\omega}\right\rangle  & \ast\\
0 & \tilde{\Delta}_{\alpha}\left\vert \tilde{\omega}\right\rangle
\end{array}
\right)
\]
with $\tilde{\Delta}_{\alpha}=\prod_{n=1}^{M}a\left(  \alpha-\beta_{n}\right)
$. We can act on $\left\vert \omega\right\rangle _{2}$ with the `combined'
monodromy matrix:
\[
\Xi_{\alpha}^{\left(  2\right)  }\left\vert \omega\right\rangle _{2}=\left(
\begin{array}
[c]{cc}%
\Delta_{\alpha}^{^{\prime\prime}}\left\vert \omega\right\rangle _{2} & \ast\\
0 & \tilde{\Delta}_{\alpha}\Delta_{\alpha}^{^{\prime\prime\prime}}\left\vert
\omega\right\rangle _{2}%
\end{array}
\right)  ~\text{.}%
\]
Choosing the reference state $\left\vert \tilde{\omega}\right\rangle $ is
equivalent to considering only $B^{2}$'s and no $B^{3}$'s when building
eigenstates. This is the trivial case when the electrons do not hybridize with
the impurity (we want instead to include non-trivial eigenstates involving
some $B^{3}$'s as well). In the more general case that concerns us we write
instead (inspired by the form of the \textit{unwanted terms}):
\begin{multline*}
\left\vert \vec{\beta},\vec{\gamma}\right\rangle =\sum_{\left\{
s_{i}\right\}  }B_{\beta_{1}}^{s_{1}}\ldots B_{\beta_{M}}^{s_{M}}\left[
B_{\gamma_{1}}^{\left(  2\right)  }\ldots B_{\gamma_{M^{\prime}}}^{\left(
2\right)  }\left\vert \omega\right\rangle _{2}\right]  _{s_{1}\ldots s_{M}%
}\equiv\\
\equiv B_{\beta_{1}}^{s_{1}}\ldots B_{\beta_{M}}^{s_{M}}\left\vert \vec
{\gamma}\right\rangle ~\text{.}%
\end{multline*}
Let us denote by $t_{\alpha}$ the eigenvalue of $T\left(  \alpha\right)  $ and
by $t_{\alpha}^{\left(  2\right)  }$ the eigenvalue of $T^{\left(  2\right)
}\left(  \alpha\right)  =A_{\alpha}^{\left(  2\right)  }+D_{\alpha}^{\left(
2\right)  }$. We need to solve the auxiliary eigenvalue subproblem:
\[
T_{\alpha}^{\left(  2\right)  }\left\vert \vec{\gamma}\right\rangle
=t_{\alpha}^{\left(  2\right)  }\left\vert \vec{\gamma}\right\rangle
\]
(notice that both $T_{\alpha}^{\left(  2\right)  }$ and $\left\vert
\vec{\gamma}\right\rangle $ are, although it is not explicitly indicated,
functions of $\vec{\beta}$). Using the results from the $SU\!\left(  2\right)
$ case we write down the \textit{wanted} terms:
\begin{align*}
\left.  A_{\alpha}^{\left(  2\right)  }\left\vert \vec{\gamma}\right\rangle
\right\vert _{\text{wanted}}  &  =\Delta_{\alpha}^{\prime\prime}\left(
\prod_{p=1}^{M^{\prime}}u_{\alpha-\gamma_{p}}\right)  \left\vert \vec{\gamma
}\right\rangle \\
\left.  D_{\alpha}^{\left(  2\right)  }\left\vert \vec{\gamma}\right\rangle
\right\vert _{\text{wanted}}  &  =\tilde{\Delta}_{\alpha}\Delta_{\alpha
}^{\prime\prime\prime}\left(  \prod_{p=1}^{M^{\prime}}u_{\gamma_{p}-\alpha
}\right)  \left\vert \vec{\gamma}\right\rangle
\end{align*}
and also the generic form of the \textit{unwanted} ones:
\begin{multline*}
\left.  A_{\alpha}^{\left(  2\right)  }\left\vert \vec{\gamma}\right\rangle
\right\vert _{\text{unwanted}}=\\
=\Delta_{\gamma_{q}}^{\prime\prime}v_{\alpha-\gamma_{q}}\left(  \prod_{p\neq
q}^{M^{\prime}}u_{\gamma_{q}-\gamma_{p}}\right)  \left\vert \alpha\,\gamma
_{1}\ldots\widehat{\gamma_{q}}\ldots\gamma_{M^{\prime}}\right\rangle
\end{multline*}%
\begin{multline*}
\left.  D_{\alpha}^{\left(  2\right)  }\left\vert \vec{\gamma}\right\rangle
\right\vert _{\text{unwanted}}=\\
=\tilde{\Delta}_{\gamma_{q}}\Delta_{\gamma_{q}}^{\prime\prime\prime}%
v_{\gamma_{q}-\alpha}\left(  \prod_{p\neq q}^{M^{\prime}}u_{\gamma_{p}%
-\gamma_{q}}\right)  \left\vert \alpha\,\gamma_{1}\ldots\widehat{\gamma_{q}%
}\ldots\gamma_{M^{\prime}}\right\rangle ~\text{.}%
\end{multline*}
The cancellation of the \textit{unwanted} terms gives a first set of auxiliary
conditions:
\[
\prod_{p\neq q}^{M^{\prime}}\frac{\gamma_{q}-\gamma_{p}-ic}{\gamma_{q}%
-\gamma_{p}+ic}=\frac{\gamma_{q}-\alpha_{0}+i\frac{c}{2}}{\gamma_{q}%
-\alpha_{0}+i\frac{3}{2}c}\prod_{n=1}^{M}\frac{\gamma_{q}-\beta_{n}}%
{\gamma_{q}-\beta_{n}+ic}%
\]
\newline whereas the \textit{wanted} terms give the auxiliary eigenvalue:
\[
t_{\alpha}^{\left(  2\right)  }=\Delta_{\alpha}^{\prime\prime}\left(
\prod_{p=1}^{M^{\prime}}u_{\alpha-\gamma_{p}}\right)  +\Delta_{\alpha}%
^{\prime\prime\prime}\tilde{\Delta}_{\alpha}\left(  \prod_{p=1}^{M^{\prime}%
}u_{\gamma_{p}-\alpha}\right)
\]
(notice that, since $\Delta_{\alpha_{j}}^{\prime\prime}=0$, the auxiliary
eigenvalues vanish, \textit{i.e.} $t_{\alpha_{j}}^{\left(  2\right)  }=0$).

Having solved the auxiliary nested problem we can go back to the original
eigenvalue problem that we are trying to solve. The combined \textit{wanted}
terms are:%
\begin{widetext}%
\[
\left.  T_{\alpha}\left\vert \vec{\beta},\vec{\gamma}\right\rangle \right\vert
_{\text{wanted}}=B_{\beta_{1}}^{s_{1}}\ldots B_{\beta_{M}}^{s_{M}}\;\left[
\left(  \prod_{n=1}^{M}u_{\alpha-\beta_{n}}\right)  +\left(  \prod_{n=1}%
^{M}u_{\beta_{n}-\alpha}\right)  t_{\alpha}^{\left(  2\right)  }\right]
\;\left\vert \vec{\gamma}\right\rangle =t_{\alpha}\left\vert \vec{\beta}%
,\vec{\gamma}\right\rangle ~\text{.}%
\]
And we find that the eigenvalues of $Z_{j}=T_{\alpha_{j}}$ read as
\[
z_{j}=t_{\alpha_{j}}=\prod_{n=1}^{M}u_{\alpha_{j}-\beta_{n}}=\prod_{n=1}%
^{M}\frac{\alpha_{j}-\beta_{n}-ic}{\alpha_{j}-\beta_{n}}~\text{.}%
\]
We are only left with the task of taking care of the \textit{unwanted} terms.
Generic ones read:
\begin{multline*}
\left.  A_{\alpha}\left\vert \vec{\beta},\vec{\gamma}\right\rangle \right\vert
_{\text{unwanted}}=A_{\alpha}\left(  B_{\beta_{n}}^{s_{n}}B_{\beta_{1}}%
^{s_{1}}\ldots\widehat{B_{\beta_{n}}^{s_{n}}}\ldots B_{\beta_{M}}^{s_{M}%
}\right)  \left(  S_{s_{1}s_{n}}^{\left(  2\right)  }\ldots S_{s_{n-1}s_{n}%
}^{\left(  2\right)  }\right)  \left\vert \vec{\gamma}\right\rangle \\
=v_{\alpha-\beta_{n}}\left(  \prod_{m\neq n}^{M}u_{\beta_{n}-\beta_{m}%
}\right)  \left(  B_{\alpha}^{s_{n}}B_{\beta_{1}}^{s_{1}}\ldots\widehat
{B_{\beta_{n}}^{s_{n}}}\ldots B_{\beta_{M}}^{s_{M}}\right)  \left(
S_{s_{1}s_{n}}^{\left(  2\right)  }\ldots S_{s_{n-1}s_{n}}^{\left(  2\right)
}\right)  \left\vert \vec{\gamma}\right\rangle
\end{multline*}
and
\begin{multline*}
\left.  D_{\alpha}\left\vert \vec{\beta},\vec{\gamma}\right\rangle \right\vert
_{\text{unwanted}}=D_{\alpha}\left(  B_{\beta_{n}}^{s_{n}}B_{\beta_{1}}%
^{s_{1}}\ldots\widehat{B_{\beta_{n}}^{s_{n}}}\ldots B_{\beta_{M}}^{s_{M}%
}\right)  \left(  S_{s_{1}s_{n}}^{\left(  2\right)  }\ldots S_{s_{n-1}s_{n}%
}^{\left(  2\right)  }\right)  \left\vert \vec{\gamma}\right\rangle \\
=v_{\beta_{n}-\alpha}\left(  \prod_{m\neq n}^{M}u_{\beta_{m}-\beta_{n}%
}\right)  \left(  B_{\alpha}^{s_{n}}B_{\beta_{1}}^{s_{1}}\ldots\widehat
{B_{\beta_{n}}^{s_{n}}}\ldots B_{\beta_{M}}^{s_{M}}\right)  \left(
S_{s_{1}s_{n}}^{\left(  2\right)  }\ldots S_{s_{n-1}s_{n}}^{\left(  2\right)
}\right)  D_{\beta_{n}}\Xi_{\beta_{n}}^{\left(  2\right)  }\left\vert
\vec{\gamma}\right\rangle
\end{multline*}
so that
\[
\left.  \operatorname*{tr}\nolimits_{A}D_{\alpha}\left\vert \vec{\beta}%
,\vec{\gamma}\right\rangle \right\vert _{\text{unwanted}}=t_{\beta_{n}%
}^{\left(  2\right)  }v_{\beta_{n}-\alpha}\left(  \prod_{m\neq n}^{M}%
u_{\beta_{m}-\beta_{n}}\right)  \left(  B_{\alpha}^{s_{n}}B_{\beta_{1}}%
^{s_{1}}\ldots\widehat{B_{\beta_{n}}^{s_{n}}}\ldots B_{\beta_{M}}^{s_{M}%
}\right)  \left(  S_{s_{1}s_{n}}^{\left(  2\right)  }\ldots S_{s_{n-1}s_{n}%
}^{\left(  2\right)  }\right)  \left\vert \vec{\gamma}\right\rangle ~\text{.}%
\]%
\end{widetext}%
Since $\tilde{\Delta}_{\beta_{n}}=0$, we have the following expression for the
eigenvalue of the auxiliary problem: $t_{\beta_{n}}^{\left(  2\right)
}=\Delta_{\beta_{n}}^{\prime\prime}\left(  \prod_{p=1}^{M^{\prime}}%
u_{\beta_{n}-\gamma_{p}}\right)  $. Combining the two contributions to the
same \textit{unwanted} term and asking that it should vanish, we find a second
set of auxiliary conditions:
\[
\prod_{m\neq n}^{M}\frac{\beta_{n}-\beta_{m}-ic}{\beta_{n}-\beta_{m}+ic}%
=\prod_{j=1}^{N}\frac{\beta_{n}-\alpha_{j}}{\beta_{n}-\alpha_{j}+ic}%
\prod_{p=1}^{M^{\prime}}\frac{\beta_{n}-\gamma_{p}-ic}{\beta_{n}-\gamma_{p}%
}~\text{.}%
\]

\subsubsection{Bethe-Ansatz Equations}

Let us collect the different equations, rearrange them and highlight the final
result. Recalling that $z_{j}=e^{-ik_{j}L}$ and performing the standard shift
$\beta_{n}=\Lambda_{n}^{\left(  1\right)  }-i\frac{c}{2}$, we write the
eigenvalue equation:
\[
e^{ik_{j}L}=\prod_{n=1}^{M_{1}}\frac{\alpha_{j}-\Lambda_{n}^{\left(  1\right)
}+i\frac{c}{2}}{\alpha_{j}-\Lambda_{n}^{\left(  1\right)  }-i\frac{c}{2}}%
\]
and after shifting $\gamma_{n}=\Lambda_{n}^{\left(  2\right)  }-ic$, we write
down also the auxiliary conditions:
\begin{multline*}
\prod_{m\neq n}^{M_{1}}\frac{\Lambda_{n}^{\left(  1\right)  }-\Lambda
_{m}^{\left(  1\right)  }-ic}{\Lambda_{n}^{\left(  1\right)  }-\Lambda
_{m}^{\left(  1\right)  }+ic}=\prod_{j=1}^{N}\frac{\Lambda_{n}^{\left(
1\right)  }-\alpha_{j}-i\frac{c}{2}}{\Lambda_{n}^{\left(  1\right)  }%
-\alpha_{j}+i\frac{c}{2}}\times\\
\times\prod_{m=1}^{M_{2}}\frac{\Lambda_{n}^{\left(  1\right)  }-\Lambda
_{m}^{\left(  2\right)  }-i\frac{c}{2}}{\Lambda_{n}^{\left(  1\right)
}-\Lambda_{m}^{\left(  2\right)  }+i\frac{c}{2}}%
\end{multline*}
and%
\begin{multline*}
\prod_{m\neq n}^{M_{2}}\frac{\Lambda_{n}^{\left(  2\right)  }-\Lambda
_{m}^{\left(  2\right)  }-ic}{\Lambda_{n}^{\left(  2\right)  }-\Lambda
_{m}^{\left(  2\right)  }+ic}=\frac{\Lambda_{n}^{\left(  2\right)  }%
-\alpha_{0}-i\frac{c}{2}}{\Lambda_{n}^{\left(  2\right)  }-\alpha_{0}%
+i\frac{c}{2}}\times\\
\times\prod_{n=1}^{M_{1}}\frac{\Lambda_{n}^{\left(  2\right)  }-\Lambda
_{n}^{\left(  1\right)  }-i\frac{c}{2}}{\Lambda_{n}^{\left(  2\right)
}-\Lambda_{n}^{\left(  1\right)  }+i\frac{c}{2}}%
\end{multline*}
where we have taken $M_{1}=M$ and $M_{2}=M^{\prime}$.

\subsection{Generalization to Models with $SU\!\left(  N\right)  $ Symmetry}

It is straightforward to generalize these results and write down the
Bethe-Ansatz equations for the more general case of $SU\!\left(  N\right)  $
internal symmetry. We use the notation $M_{0}=N_{\mathrm{e}}$ for the number
of electrons and $M_{N}=N_{\mathrm{i}}\;\left(  =1\right)  $ for the
\textit{number of impurities}. We define $\Lambda_{n}^{\left(  0\right)
}=\alpha_{n}\;\left(  =k_{n}\right)  $ for the charge rapidities and
$\Lambda_{n}^{\left(  N\right)  }=\varepsilon_{n}\;\left(  =\varepsilon
\right)  $ (\textit{i.e.}~the \textit{impurity rapidities}). Then we write the
eigenvalue equations:
\[
e^{ik_{n}L}=\prod_{m=1}^{M_{1}}\frac{\Lambda_{n}^{\left(  0\right)  }%
-\Lambda_{m}^{\left(  1\right)  }+i\frac{c}{2}}{\Lambda_{n}^{\left(  0\right)
}-\Lambda_{m}^{\left(  1\right)  }-i\frac{c}{2}}%
\]
and the nested auxiliary conditions:
\[
\prod_{m\neq n}^{M_{r}}\frac{\Lambda_{n}^{\left(  r\right)  }-\Lambda
_{m}^{\left(  r\right)  }-ic}{\Lambda_{n}^{\left(  r\right)  }-\Lambda
_{m}^{\left(  r\right)  }+ic}=\prod_{\sigma=\pm1}\prod_{m=1}^{M_{r+\sigma}%
}\frac{\Lambda_{n}^{\left(  r\right)  }-\Lambda_{m}^{\left(  r+\sigma\right)
}-i\frac{c}{2}}{\Lambda_{n}^{\left(  r\right)  }-\Lambda_{m}^{\left(
r+\sigma\right)  }+i\frac{c}{2}}%
\]
where the \textit{rank} varies in the range $r=1,\ldots,N-1$.

\subsubsection{Generalization to Models with $SU\!\left(  N\right)  \otimes
SU\!\left(  M\right)  $ Symmetry}

We start with the same auxiliary eigenvalue problem, but with the scattering
matrices:
\begin{align*}
S_{jn\neq0}  &  =\frac{\left(  \alpha_{j}-\alpha_{n}\right)  \mathbf{I}%
_{jn}^{s}-ic\mathbf{P}_{jn}^{s}}{\left(  \alpha_{j}-\alpha_{n}\right)
-ic}\frac{\left(  \alpha_{j}-\alpha_{n}\right)  \mathbf{I}_{jn}^{q}%
+ic\mathbf{P}_{jn}^{q}}{\left(  \alpha_{j}-\alpha_{n}\right)  +ic}\\
S_{j0}  &  =\mathbf{I}_{j0}^{q}+\frac{e^{-i\delta\left(  \alpha_{j}-\alpha
_{0}\right)  }-1}{M}\mathbf{Q}_{j0}^{q}~\text{.}%
\end{align*}
We choose to consider a case where $S_{j0}$ acts non-trivially in the
`channel' degrees of freedom only (\textit{i.e.}~in `q-flavor' space). This
problem amounts to taking the one discussed above and adding an extra
`isospin' to it. The monodromy matrix can be written as a direct product:
$\Xi\left(  \alpha\right)  =\Xi^{s}\left(  \alpha\right)  \otimes\Xi
^{q}\left(  \alpha\right)  $, and the transfer matrix becomes
\[
T_{\alpha}=T_{\alpha}^{s}T_{\alpha}^{c}=\left(  A_{\alpha}^{s}+D_{\alpha}%
^{s}\right)  \left(  A_{\alpha}^{q}+D_{\alpha}^{q}\right)  ~\text{.}%
\]
The different steps go through as before and we get the Bethe-Ansatz Equations
(BAE):
\[
e^{ik_{j}L}=\prod_{n=1}^{M_{1}^{s}}\frac{\alpha_{j}-\Lambda_{n}^{s\left(
1\right)  }-i\frac{c}{2}}{\alpha_{j}-\Lambda_{n}^{s\left(  1\right)  }%
+i\frac{c}{2}}\prod_{m=1}^{M_{1}^{q}}\frac{\alpha_{j}-\Lambda_{m}^{q\left(
1\right)  }+i\frac{c}{2}}{\alpha_{j}-\Lambda_{m}^{q\left(  1\right)  }%
-i\frac{c}{2}}%
\]
with the conditions,
\begin{align*}
\prod_{m\neq n}^{M_{r}^{s}}\frac{\Lambda_{n}^{s\left(  r\right)  }-\Lambda
_{m}^{s\left(  r\right)  }-ic}{\Lambda_{n}^{s\left(  r\right)  }-\Lambda
_{m}^{s\left(  r\right)  }+ic}  &  =\prod_{\sigma=\pm1}\prod_{m=1}%
^{M_{r+\sigma}^{s}}\frac{\Lambda_{n}^{s\left(  r\right)  }-\Lambda
_{m}^{s\left(  r+\sigma\right)  }-i\frac{c}{2}}{\Lambda_{n}^{s\left(
r\right)  }-\Lambda_{m}^{s\left(  r+\sigma\right)  }+i\frac{c}{2}}\\
\prod_{m\neq n}^{M_{r}^{q}}\frac{\Lambda_{n}^{q\left(  r\right)  }-\Lambda
_{m}^{q\left(  r\right)  }-ic}{\Lambda_{n}^{q\left(  r\right)  }-\Lambda
_{m}^{q\left(  r\right)  }+ic}  &  =\prod_{\sigma=\pm1}\prod_{m=1}%
^{M_{r+\sigma}^{q}}\frac{\Lambda_{n}^{q\left(  r\right)  }-\Lambda
_{m}^{q\left(  r+\sigma\right)  }-i\frac{c}{2}}{\Lambda_{n}^{q\left(
r\right)  }-\Lambda_{m}^{q\left(  r+\sigma\right)  }+i\frac{c}{2}}%
\end{align*}
where for convenience we have used the definitions
\begin{align*}
\Lambda_{n}^{s,q\left(  0\right)  }  &  =\alpha_{n}\\
\Lambda_{1}^{q\left(  M\right)  }  &  =\alpha_{0}%
\end{align*}
and accordingly $M_{0}^{s,q}=N_{\mathrm{e}}$, $M_{M}^{q}=N_{\mathrm{i}}=1$,
and $M_{N}^{s}=0$. One sees that the effect of the impurity enters via the
auxiliary conditions for the \textit{q-flavor--rapidities}.

The solution thus far was general. To specify it to our model we take
$\alpha_{j}=k_{j}$ and $c=2\Delta\equiv V^{2}$, and obtain the Bethe-Ansatz
equations for the multi-channel Anderson impurity model. Removing the impurity
(\textit{i.e.}~taking $M_{M}^{q}=0$) one recovers the usual equations of the
NBA formalism.\cite{tw83}

\section{Kernels and Identities\label{kernels}}

In this appendix we collect a number of function and kernel definitions, and
identities relating them, that are central to the writing, rewriting, and
algebraic manipulation of the BAE and the equations of the TBA. Let us
introduce the following notations aimed at lightening the writing of the BAE,
\begin{align*}
e_{n}\left(  z\right)   &  =\frac{z-in\Delta}{z+in\Delta}\\
e_{nm}^{\prime}\left(  z\right)   &  =\prod_{\tau=1}^{\min\left\{
m,n\right\}  }e_{m+n+1-2\tau}\left(  z\right) \\
e_{nm}^{\prime\prime}\left(  z\right)   &  =\prod_{\tau=1}^{\min\left\{
m,n\right\}  }e_{m+n-2\tau}\left(  z\right) \\
e_{nm}\left(  z\right)   &  =\prod_{\sigma=\pm1}e_{n\,m+\sigma}^{\prime
}\left(  z\right)  ~\text{.}%
\end{align*}
For the purpose of writing a continuum version of the BAE, we will need as
well the derivatives of the logarithms of the above functions. We define the
following kernels:
\[
K_{n}\left(  z\right)  =\left(  2\pi i\right)  ^{-1}\partial_{z}\ln
e_{n}\left(  z\right)  =\frac{1}{\pi}\frac{n\Delta}{z^{2}+\left(
n\Delta\right)  ^{2}}%
\]
plus the similar definitions
\begin{align*}
K_{nm}\left(  z\right)   &  =\left(  2\pi i\right)  ^{-1}\partial_{z}\ln
e_{nm}\left(  z\right) \\
K_{nm}^{\prime}\left(  z\right)   &  =\left(  2\pi i\right)  ^{-1}\partial
_{z}\ln e_{nm}^{\prime}\left(  z\right) \\
K_{nm}^{\prime\prime}\left(  z\right)   &  =\left(  2\pi i\right)
^{-1}\partial_{z}\ln e_{nm}^{\prime\prime}\left(  z\right)  ~\text{.}%
\end{align*}
These last three definitions are the basis to define the following
`convolution' kernels:
\begin{align*}
A_{nm}\left(  z\right)   &  =\delta_{n,m}\delta\left(  z\right)
+K_{nm}\left(  z\right)  =\\
&  \qquad\qquad\qquad=\sum_{\sigma=\pm1}\sum_{\tau=1}^{\min\left\{
m,n\right\}  }K_{m+n+1+\sigma-2\tau}\left(  z\right) \\
B_{nm}\left(  z\right)   &  =K_{nm}^{\prime}\left(  z\right)  =\sum_{\tau
=1}^{\min\left\{  m,n\right\}  }K_{m+n+1-2\tau}\left(  z\right) \\
C_{nm}\left(  z\right)   &  =\delta_{n,m}\delta\left(  z\right)
+K_{nm}^{\prime\prime}\left(  z\right)  =\sum_{\tau=1}^{\min\left\{
m,n\right\}  }K_{m+n-2\tau}\left(  z\right)
\end{align*}
that are used extensively in the continuum formulation of the BAE.

\subsubsection{Fourier Space Formalism}

A great simplification in the algebraic manipulations is often achieved by
working in terms of the Fourier transformed densities. Our convention will be
as follows:
\[
\tilde{\rho}\left(  w\right)  \equiv\mathcal{F}\left\{  \rho\left(  z\right)
\right\}  \left(  w\right)  =\int\rho\left(  z\right)  \;e^{-iwz}dz~\text{.}%
\]
So that, for instance, the basic kernels adopt the simple form $\tilde{K}%
_{n}\left(  w\right)  =e^{-n\Delta\left\vert w\right\vert }$. Working in
Fourier space is easy to see that different convolution kernels are simply
related: $\tilde{B}_{n,m}=\tilde{K}_{1}\tilde{C}_{n,m}$ and $\tilde{B}%
_{n,m}=\tilde{G}\tilde{A}_{n,m}$. Where one defines the `basic recursion
kernel',
\begin{align*}
\tilde{G}\left(  w\right)   &  =\frac{\tilde{K}_{1}\left(  w\right)  }%
{\tilde{K}_{0}\left(  w\right)  +\tilde{K}_{2}\left(  w\right)  }=\frac
{1}{2\cosh\Delta w}\\
G\left(  z\right)   &  =\mathcal{F}^{-1}\left\{  \tilde{G}\left(  w\right)
\right\}  \left(  z\right)  =\frac{1}{4\Delta\cosh\frac{\pi z}{2\Delta}}%
\end{align*}
that we call so because it enters many recursion relations connecting the
different convolution kernels (for instance:\cite{Andrei} $\tilde{A}%
_{n,m}-\tilde{G}\tilde{A}_{n+1,m}+\left(  1-\delta_{n,1}\right)  \tilde
{G}\tilde{A}_{n-1,m}=\delta_{n,m}$ plus many others involving also the
convolution kernels $\tilde{B}_{n,m}$ and $\tilde{C}_{n,m}$). All these
relations are easy to prove in Fourier space.

\subsubsection{General Recursion Kernels}

The basic recursion kernel is ubiquitous in the TBA equations of all
integrable models. For the case of the multi-channel Anderson model, we are
also going to need the more general kernels:
\[
\tilde{G}_{m}^{\left(  N,M\right)  }=\frac{\tilde{A}_{Nm}}{\tilde{A}_{MM}%
}=e^{-\left(  N-M\right)  \Delta\left\vert w\right\vert }\frac{\sinh\left(
m\Delta w\right)  }{\sinh\left(  M\Delta w\right)  }~\text{.}%
\]
Of which the basic recursion kernel is a particular case, $G=G_{1}^{\left(
2,2\right)  }$. When $m=N$ some of these kernels are singular (have non-zero
asymptotics in Fourier space). We regularize them according to
\[
{}^{R}G_{m}^{\left(  N,M\right)  }\left(  \lambda\right)  =G_{m}^{\left(
N,M\right)  }\left(  \lambda\right)  -\delta_{m,N}~\delta\left(
\lambda\right)
\]
where we have expressed them, in direct space, in terms of the variable
$\lambda=\pi z/2\Delta$. When carrying out numerical calculations, we will use
the following explicit expressions for these kernels:%
\begin{multline*}
{}^{R}G_{m}^{\left(  N,M\right)  }=\\
=\frac{1}{2\pi}\frac{1}{M\pi}\sum_{\substack{\sigma=\pm1\\\tau=\pm1}%
}\tau~\digamma\left(  \delta_{N+\tau m,0}+\frac{N+\tau m}{2M}+i\frac
{\sigma\lambda}{M\pi}\right)
\end{multline*}
where $\digamma\left(  z\right)  \equiv\partial_{z}\ln\Gamma\left(  z\right)
$ is the digamma function.

\subsubsection{Rank Recursion Kernels}

We also need, for intermediate manipulations, kernel operators that act on the
rank indices. The basic one is $\tilde{G}^{rs}=\delta_{s}^{r}-\left(
\delta_{s}^{r+1}+\delta_{s}^{r-1}\right)  \tilde{G}$. Using it we can extend
the convolution kernels as $\tilde{A}_{nm}^{rs}=\tilde{G}^{rs}\tilde{A}_{nm}$.
A particularly useful derived kernel is the one given by the inverse:
$R_{X}^{rs}\equiv\left[  G^{-1}G^{sr}\right]  ^{-1}$, where the indices vary
in the range $1,\ldots,N-1$ or $1,\ldots,M-1$ depending on the case ($X=N,M$).
The explicit formula for this operator, in Fourier space, is%
\[
\tilde{R}_{X}^{rs}=\frac{\sinh\left[  \min\left(  r,s\right)  \Delta w\right]
\sinh\left[  \left(  X-\max\left(  r,s\right)  \right)  \Delta w\right]
}{\sinh\left(  \Delta w\right)  \sinh\left(  X\Delta w\right)  }%
\]
and two particularly useful cases are given by the general recursion kernels:%
\begin{align*}
R_{M}^{r1}\left(  \lambda\right)   &  =G_{M-r}^{\left(  M,M\right)  }\left(
\lambda\right)  =\frac{\frac{1}{\pi M}\sin\frac{\pi\left(  M-r\right)  }{M}%
}{\cos\frac{\pi\left(  M-r\right)  }{M}+\cosh\frac{2\lambda}{M}}\\
R_{N}^{nN-1}\left(  \lambda\right)   &  =G_{n}^{\left(  N,N\right)  }\left(
\lambda\right)  \ =\frac{\frac{1}{\pi N}\sin\frac{\pi n}{N}}{\cos\frac{\pi
n}{N}+\cosh\frac{2\lambda}{N}}%
\end{align*}
(these will be used in the determination of the Schrieffer-Wolff limit).


\end{document}